\documentclass{article}

\usepackage{arxiv}
\usepackage[utf8]{inputenc} 
\usepackage[T1]{fontenc}    
\usepackage{url}            
\usepackage{booktabs}       
\usepackage{amsmath, amsfonts, amssymb}       
\usepackage{nicefrac}       
\usepackage{microtype}      
\usepackage{lipsum}
\usepackage{graphicx}
\usepackage[bottom]{footmisc}
\usepackage[english]{babel}
\usepackage{caption}
\usepackage[normalem]{ulem}
\usepackage{color,soul}
\usepackage{authblk}
\usepackage{svg}
\usepackage{enumitem}
\usepackage{subcaption}
\newcommand{\beginsupplement}{%
        \setcounter{table}{0}
        \renewcommand{\thetable}{S\arabic{table}}%
        \setcounter{figure}{0}
        \renewcommand{\thefigure}{S\arabic{figure}}%
        \setcounter{section}{0}
        \renewcommand{\thesection}{S\arabic{section}}%
     }

\usepackage{caption,graphicx,newfloat}
\usepackage[font=scriptsize,labelfont=bf]{caption}
\DeclareCaptionType{InfoBox}

\usepackage{verbatim}
\usepackage{algorithmic}
\usepackage{multicol, blindtext}
\usepackage{textcomp}

\def\BibTeX{{\rm B\kern-.05em{\sc i\kern-.025em b}\kern-.08em
    T\kern-.1667em\lower.7ex\hbox{E}\kern-.125emX}}

\usepackage{lmodern}
\usepackage[T1]{fontenc}
\usepackage{float}

\usepackage{hyperref, url}
\usepackage{xr}

\makeatletter
\newcommand*{\addFileDependency}[1]{
  \typeout{(#1)}
  \@addtofilelist{#1}
  \IfFileExists{#1}{}{\typeout{No file #1.}}
}
\makeatother

\begin{document}

{
\textbf\newline \title{\LARGE Unavailability of experimental 3D structural data on protein folding dynamics and necessity for a new generation of structure prediction methods in this context$^\dagger$}
}

\vspace{-1cm}

\author{Aydin Wells$^{1}$, Khalique Newaz$^{2}$, Jennifer Morones$^{1}$, Jianlin Cheng$^{3}$, and Tijana Milenkovi\'{c}$^{1,*}$}
\date{}

\maketitle
\begingroup
\renewcommand\thefootnote{$\dagger$}
\footnotetext{This manuscript has been published in Bioinformatics. DOI:  10.1093/bioinformatics/btag020}
\endgroup

\begin{center}
\vspace{-0.8cm}

$^1$Department of Computer Science and Engineering, University of Notre Dame, USA\\
$^2$Institute for Computational Systems Biology, University of Hamburg, Germany\\ 
$^3$Department of Electrical Engineering and Computer Science, University of Missouri, Columbia, USA\\
$^*$Corresponding author (email: tmilenko@nd.edu; mailing address: 381 Fitzpatrick Hall of Engineering, University of Notre Dame, Notre Dame, IN 46556).\\

\end{center}

\bigskip

\begin{abstract}
\textbf{Motivation: }Protein folding is a dynamic process during which a protein's amino acid sequence undergoes a series of 3-dimensional (3D) conformational changes en route to reaching a native 3D structure; these conformations are called folding intermediates. While data on native 3D structures are abundant, data on 3D structures of non-native intermediates remain sparse, due to limitations of current  technologies for experimental determination of 3D structures. Yet, analyzing folding intermediates is crucial for understanding folding dynamics and misfolding-related diseases. Hence, we search the literature for available (experimentally and computationally obtained) 3D structural data  on folding intermediates, organizing the data in a centralized resource. Also, we assess whether existing methods, designed for predicting native structures, can also be utilized to predict structures of non-native intermediates.
\\ 
\textbf{Results: }Our literature search reveals six studies that provide 3D structural data on folding intermediates (two for post-translational and four for co-translational folding), each focused on a single protein, with 2-4 intermediates. Our assessment shows that an established method for predicting native structures, AlphaFold2, does not perform well for non-native intermediates in the context of co-translational folding; a recent study on post-translational folding concluded the same for even more existing methods. Yet, we identify in the literature recent pioneering methods designed explicitly to predict 3D structures of folding intermediates by incorporating intrinsic biophysical characteristics of folding dynamics, which show promise. This study assesses the current landscape and future directions of the field of 3D structural analysis of protein folding dynamics.
\\
\textbf{Availability and implementation: }\url{https://github.com/Aywells/3Dpfi} or \url{https://www3.nd.edu/~cone/3Dpfi}
\end{abstract}
\section{Introduction}\label{sect:introduction}

    Protein folding is a process by which a protein's amino acid sequence folds into a three-dimensional (3D) structure (or conformation). The 3D structure directs what other biomolecules the protein may interact with to carry out its function(s) \cite{alberts2002shape, gething1992protein}. When a protein folds incorrectly, its misfolded 3D structure can disrupt normal cellular functions and contribute to diseases. Protein folding is a dynamic process, as a protein's sequence gradually folds into a series of 3D structural conformations until arriving at a final, native structure \cite{privalov1996intermediate, wang2025cotranslational}; this time series of 3D structural changes (including the native state) is called a \emph{protein folding pathway}. The dynamics of the folding process can be studied from two prominent perspectives: \textit{post-}  and \textit{ co-translational} folding. 
    
\begin{figure*}
        \begin{center}
        \includegraphics[width=0.94\textwidth,trim= 0.2cm 16.2cm 0cm 2.5cm]{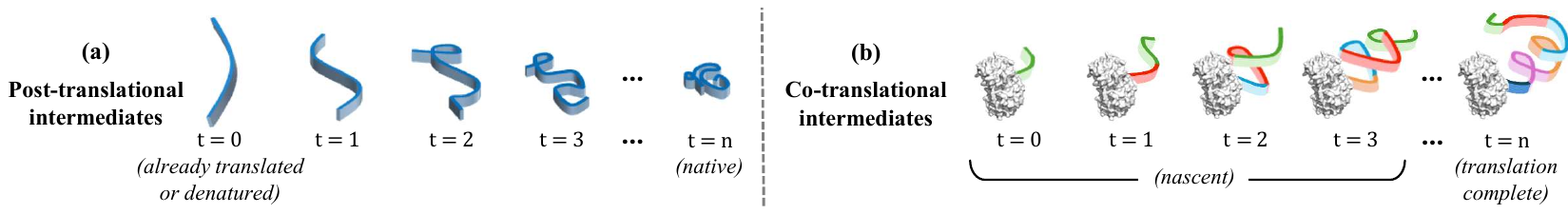}
        \end{center}
        \caption{Illustrations of \textbf{(a)} post- and \textbf{(b)} co-translational protein folding pathway intermediates. \textbf{(a)} A protein's \textit{entire} sequence (blue line) undergoes conformational changes. The resulting 3D structures from time 0 to time $n$ are the intermediates in the post-translational pathway. \textbf{(b)} A protein's nascent sequence undergoes conformational changes as newly translated amino acids are added (additional line color) by the ribosome (grey entity) at time $t=k$ to its already translated part from the previous time step ($t=k-1$), until all amino acids have been translated ($t=n$). The resulting 3D structures are the intermediates in the co-translational pathway.  In both panels, the intermediates from time $0$ to time $n-1$ are non-native, and the  intermediate at time $n$ is native. Note that the notion of an intermediate is tied to time, and not to the portion of sequence present in the 3D fold at time $t$:  regardless of whether the 3D structure at time $t$ is for the full sequence (panel (a)) or a subsequence (panel (b)), the structure is considered to be a (post- or co-translational, respectively) intermediate at time $t$. Also, note that both a post- and co-translational non-native intermediate can be \emph{on-pathway}, meaning one that is traversed to reach the native state, or \emph{off-pathway}, corresponding to a misfolded state that must be resolved before correct folding can continue \cite{baldwin1996pathway, clark2004protein}.
        This means that ``an intermediate'' can be any (on- or off-pathway) intermediate on a (mis)folding pathway, rather than exclusively an intermediate with distinct biophysical characterizations such as local minima on the folding energy landscape. In our paper, intermediates are whatever 3D structures are provided along a folding pathway in considered data.
        %
        %
        }
        \label{fig:pathway_n_cotrans}
    \end{figure*}

    Post-translational folding (aka folding in vitro \cite{duran2025native}, folding in solution \cite{wang2025cotranslational}, or free folding \cite{tao2025pathway}) is concerned with the structural changes of a protein's \emph{entire} sequence, which may be already translated (defined below) or instead may start from a denatured state \cite{basharov2003protein}. Specifically, a post-translational protein folding pathway is a time-series of 3D conformational changes that a protein's full sequence undergoes until reaching a native state (Fig. \ref{fig:pathway_n_cotrans}(a)) \cite{basharov2003protein,englander2014nature}. The full-sequence 3D conformations formed during this process (including the native structure) are called post-translational protein folding pathway intermediates (or just post-translational intermediates, for simplicity) \cite{plessa2021nascent}. 

    Co-translational folding is concerned with the structural changes of a protein's \emph{partial} sequence as it gradually grows \emph{during} its translation by the ribosome. Namely, a co-translational protein folding pathway is a time-series of conformational changes that a protein's sequence undergoes as newly translated amino acids are added to its already translated part from the previous time step, until all amino acids have been translated (Fig. \ref{fig:pathway_n_cotrans}(b)) \cite{moss2024effects,nilsson2015cotranslational, wang2025cotranslational}. The 3D conformations corresponding to the gradually increasing subsequences formed during translation (including the native structure) are called co-translational protein folding pathway intermediates (or just co-translational intermediates) \cite{jacobs2017evidence}. 

A recent study summarizes (dis)advantages of post- vs. co-translational folding \cite{duran2025native}. Briefly, most of  understanding of protein folding comes from the former, with one of the most valuable insights being that a protein's amino acid sequence encodes the information about its native 3D structure. However, even in vitro, a large fraction of the proteome cannot refold from a denatured
state and instead tends to misfold and aggregate. Moreover, in vivo, an estimated one-third of the proteome in \emph{E. coli} (as a representative prokaryote) folds at least one entire domain co-translationally, and this fraction is likely even higher in eukaryotes given their slower translation rates. Moreover, slowing down or speeding up translation rates (e.g., as approximated by codon usage) can allow an already synthesized N-terminal portion more time to fold independently from a C-terminal portion or prevent specific folding
intermediates from being populated, respectively. Consequently, the vectorial nature and translation rates are exploited in vivo to increase folding efficiency (i.e. how quickly a protein becomes functional \cite{siller2010slowing,tsytlonok2013s, wagner1999intermediates}), which has implications for why many proteins fold more efficiently co-translationally than post-translationally \cite{duran2025native}.
    
    Analyzing folding intermediates is crucial for understanding protein folding dynamics and deepening insights on protein functions and  misfolding-related diseases. 
    Some data types -- specifically kinetics and thermodynamics data -- do exist on a somewhat large scale that aim to capture information on folding dynamics. For post-translational folding, databases with  kinetics (protein folding rate) data exist for dozens to hundreds of proteins. For example, the Protein Folding DataBase (PFDB) contains kinetics (folding rate constant) data for 141 (89 two-state and 52 multi-state) single-domain globular proteins; a two-state protein folds via a two-state mechanism, directly from unfolded to folded, and a multi-state protein folds via a multi-state mechanism involving more than two intermediates.
    As of its publication in 2019, this made PFDB the largest available database of protein folding kinetics \cite{manavalan2019pfdb}. As another  example, Start2Fold contains kinetics (hydrogen/deuterium exchange) data extracted from the literature for 57 proteins with 219 residue sets (i.e. conformations), where each protein may have several residue sets categorized based on so-called protection levels (i.e. early, intermediate, and late for folding; and strong, medium, and weak for stability) \cite{pancsa2016start2fold}. For post-translational folding,  thermodynamics data also exist on a large scale. For example, ProThermDB contains thermodynamics (e.g. melting temperature and free energy) data for 31,580 (12,050 wild-type and 19,530 mutated) proteins \cite{nikam2021prothermdb}. 
    For co-translational folding, we could not identify any organized database containing either kinetics or thermodynamics data. Instead, we could identify some isolated studies that provide data of these types for a handful of proteins (Supplementary Section \ref{sect:sup_thermo}).
    
    \begin{figure*}
        \begin{center}
        \includegraphics[width=1.0\textwidth,trim= 0cm 15.5cm 0cm 0.5cm]{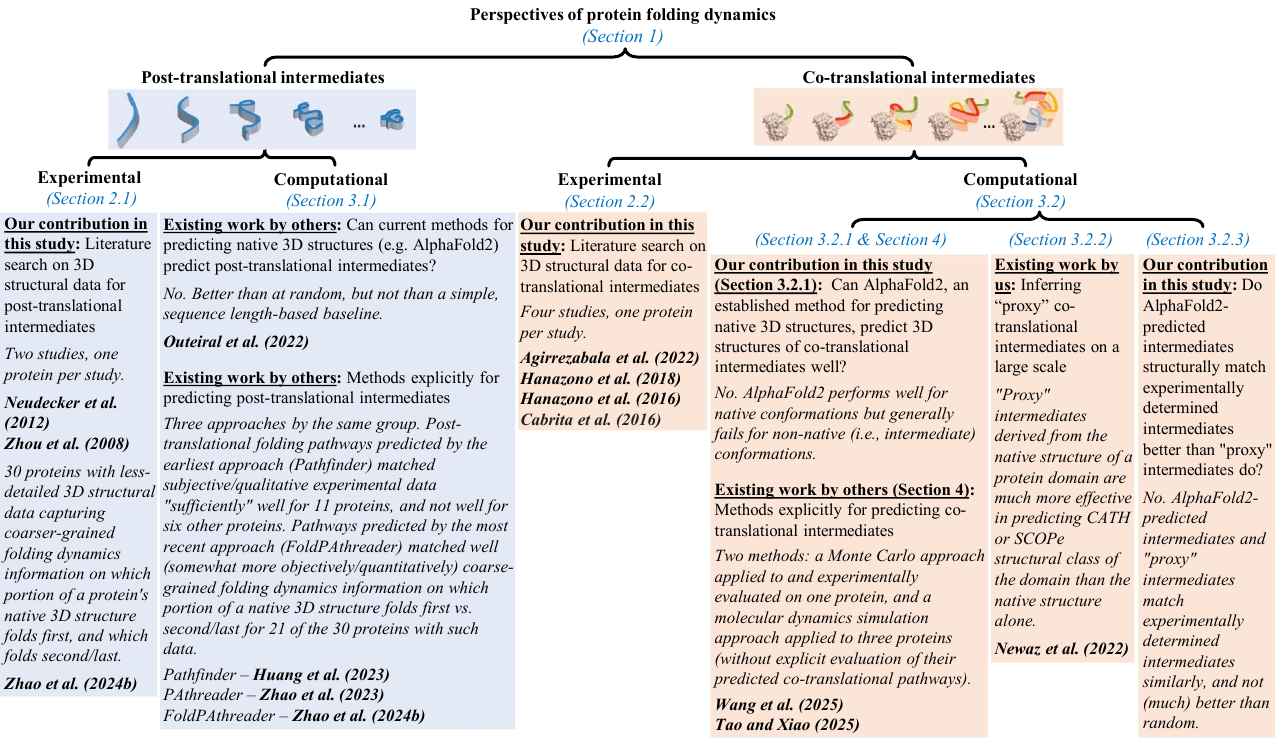}
        \end{center}
        \caption{Summary of this paper that is focused on availability and analysis of 3D structural data  related to protein folding dynamics.}
        %
        %
        \label{fig:outline}
    \end{figure*}
    
    Protein folding kinetics and thermodynamics data are valuable for understanding folding dynamics \cite{gelman2014fast, zhao2013nature}. 
    However, both provide quantitative measures of the protein folding process, and typically, a quantity per protein rather than per intermediate of a protein.  In other words, they lack 3D spatial resolution at the level of intermediates to reveal the atomic-level interactions and structural transitions that underpin the protein folding process. Similar holds for other types of data capturing properties of folding intermediates/pathways/dynamics that are obtained through HDX, site-resolved NMR spectroscopy, and mass spectrometry biophysical technologiesc \cite{englander1992protein, narang2020hdx}. HDX probes changes in protein backbone exposure to infer which regions of the protein become transiently protected or exposed during folding. Site-resolved NMR provides residue-level information on local chemical environments and conformational exchange to detect partially folded states. Mass spectrometry, often coupled with HDX, detects mass shifts associated with hydrogen exchange, thus reports on the structural stability and kinetics of intermediate states.

    To understand folding dynamics better, 3D structures of folding intermediates should be studied. Our goal is to provide a centralized resource discussing current data of this type. We comment on the availability of 3D structures of intermediates obtained (i) experimentally and (ii) computationally.
    
    First, biophysical technologies such as 
    NMR spectroscopy, cryo-EM, and X-ray crystallography \cite{klebe2025experimental} have contributed a wealth of experimentally determined 3D structures that are publicly available in the Protein Data Bank (PDB) \cite{wwpdb2019protein}. These data are almost exclusively for native states of proteins rather than for non-native folding intermediates. This is because NMR and cryo-EM may capture instances of intermediate states, but their resolution or sensitivity may fall short when trying to detect a rapidly fluctuating sequence of conformational changes \cite{klebe2025experimental}; and X-ray crystallography requires proteins to form crystals, a challenge that folding intermediates often cannot meet due to their unstable and dynamic nature \cite{klebe2025experimental, zheng2015x}. Moreover, the protein folding process (folding of secondary and tertiary structures  rather than necessarily the complete protein synthesis) often occurs in microseconds to seconds \cite{duran2025native,mayor2000protein}, making it difficult for the technologies to achieve the temporal resolution necessary to capture 3D structures of intermediates in real time \cite{freddolino2010challenges,lindorff2011fast}.
    
    Second, the availability of large-scale experimentally derived data on native 3D structures has facilitated the development of computational approaches for predicting native 3D structures on an even larger scale \cite{abramson2024accurate, Outeiral2022CurrentSP, yuan2025casp16}. As we will show, there is a lack of large-scale experimentally determined 3D structural data on non-native folding intermediates. Given this, an interesting question arises: can computational approaches for native 3D structure prediction be used to predict 3D structures of non-native folding intermediates?   
    For post-translational intermediates, this question was already asked recently, revealing a lack of success when using eight prominent approaches of this type, including AlphaFold2 \cite{Outeiral2022CurrentSP}. More recently, initial computational approaches have been introduced specifically for predicting 3D structures of post-translational intermediates \cite{Huang2023PathfinderPF,zhao2023protein,zhao2024foldpathreader}; a recent review of these approaches is available \cite{zhao2024recent}. However, these valuable attempts have focused on \emph{post-translational} intermediates. Our literature search done as a part of this study prior to the submission of our paper has found no attempts focused on predicting 3D structures of \emph{co-translational} intermediates. Hence, as a contribution of this study, we introduce an original research evaluation of whether AlphaFold2 can correctly predict experimentally determined 3D structures of co-translational intermediates that are currently available for a handful of proteins. Note that after the submission and during the revision of our paper, two relevant studies were published that used an existing Monte Carlo approach and a molecular dynamics approach tailored to predict co-translational pathways for one and three proteins, respectively \cite{wang2025cotranslational,tao2025pathway}. However, none of them evaluated their predicted pathways by directly comparing 3D structures of their predicted co-translational intermediates against 3D structures of experimentally determined intermediates.

    Unlike the existing work, in this paper  (Fig. \ref{fig:outline}), we  search the literature for available experimentally determined (Section \ref{sect:exp_data}) and computational (Section \ref{sect:comp_data}) 3D structural data on folding intermediates, both post-translational (Sections \ref{subsect:exp_pathway} and \ref{subsect:comp_pathway}) and co-translational (Sections \ref{subsect:exp_cotrans} and \ref{subsect:comp_cotrans}) ones. Also, we investigate whether current native 3D structure predictors can  predict well 3D structures of post-translational intermediates (from existing literature; Section \ref{subsect:comp_pathway}) and co-translational intermediates (from our own analysis original to this paper; Section \ref{subsect:comp_cotrans}). We aim to inform the  community about available 3D structural data on folding intermediates and a (potential) need for a new generation of protein 3D structure predictors in this context. With our coverage of 3D structural data and structure prediction methods in the context of co-translational folding, we complement the above mentioned review that is on post-translational folding \cite{zhao2024recent}.
    
Other approach families for studying folding dynamics exist: molecular dynamics and Monte Carlo simulations, and coarse-grained models \cite{kmiecik2016coarse}. We touch on these in Section \ref{subsect:comp_pathway}.

\section{Experimental 3D structural data on protein folding pathway intermediates}\label{sect:exp_data}

    \subsection{Experimental post-translational intermediates}\label{subsect:exp_pathway}

\begin{figure*}[ht]
    \begin{center}
            \includegraphics[
                width=0.95\textwidth,
                trim= 1.2cm 9cm 4.1cm 2cm,
                clip
            ]{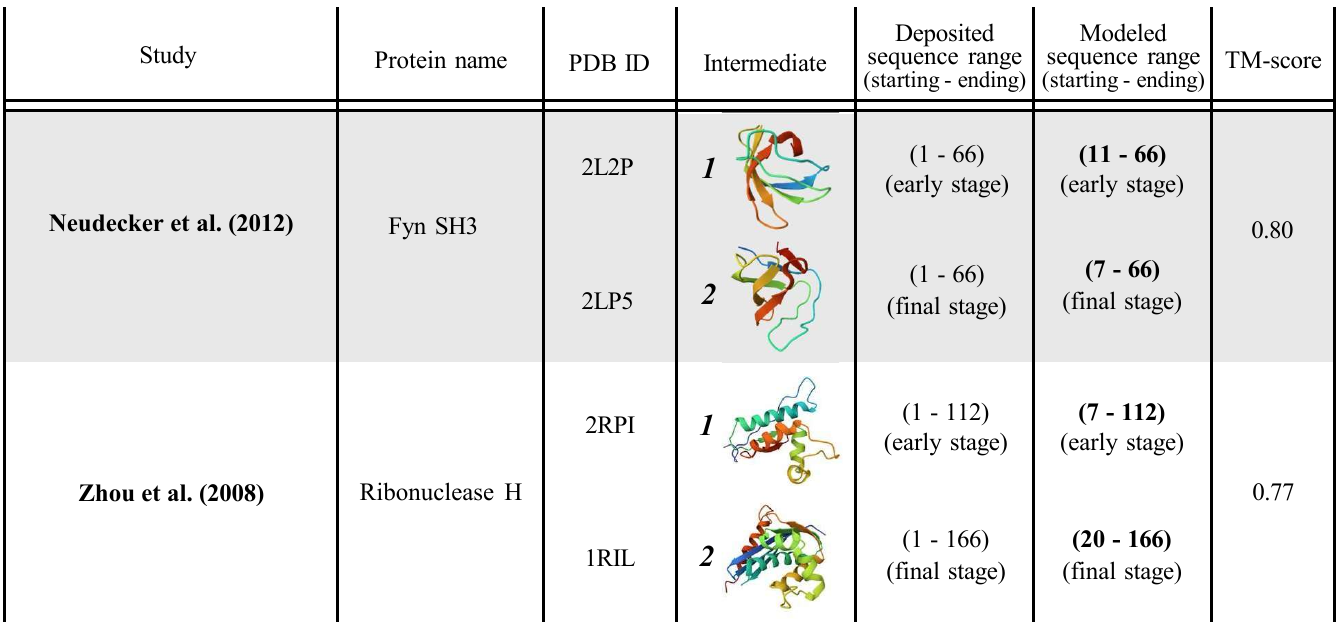}%
    \end{center}
    \caption{Details of the two studies we have identified that report 3D structural data of post-translational intermediates. For each study, i.e. its considered protein, two PDB IDs and two visualizations are shown for the protein's two intermediates (1: ``early-stage'' or pre-native state; 2: ``final stage'' or native state). In the fourth and fifth columns, the deposited sequence range and the modeled sequence range for each intermediate is provided, as reported in its respective study and PDB. Note that a bolded model sequence range value indicates that there is a range discrepancy between the reported deposited sequence and the reported model sequence. In the last column, we measure and show the structural similarity between the modeled sequence of the first intermediate and the modeled sequence of the second intermediate using TM-score \cite{zhang2004scoring}.}
    \label{fig:pathway_data}
\end{figure*}

        Our literature search on experimentally determined 3D structural data for post-translational intermediates identifies two studies \cite{neudecker2012structure,zhou2008high} that explicitly documented 3D structures of intermediates in PDB (Supplementary Fig. \ref{fig:pathway_data} and Supplementary Section \ref{sect:sup_post_studies}). Each of the two studies investigated one protein. The two proteins have sequence lengths of on the order of 100 amino acids. Each study reported  two post-translational intermediates per protein -- one  pre-native  and one  native. For each intermediate, a deposited sequence range (i.e. the amino acid sequence that is reported) and a modeled sequence range (i.e. the portion of the sequence that has available 3D structure information) were reported in PDB; note that the two can differ. 

        While these two studies have successfully resolved 3D structures of intermediates, they are limited in scope -- the number of proteins studied (only two), the size of proteins studied (quite short), and the number of intermediates reported (only two). As the current 3D structural data on post-translational intermediates is sparse, owing to the limitations of biophysical technological  discussed above, there is a need for novel  technologies for this purpose. 

        For each study's protein, we examine the 3D structural similarity between the given protein's first intermediate and its second intermediate, in terms of the Template Modeling score (TM-score) \cite{zhang2004scoring}. TM-scores range between 0 and 1, where scores below 0.17 indicate random-like similarities, scores above 0.3 indicate significant similarities, and scores above 0.5 indicate the same overall fold (Supplementary Section \ref{sect:sup_tm_score}). We find that the TM-score between the two intermediates is reasonably high, between $0.77-0.80$, depending on the protein (Supplementary Fig. \ref{fig:pathway_data}). Also, for completeness, we examine each study's aim in more detail; due to space constraints, we report this in Supplementary Section \ref{sect:sup_post_studies}.
        
        Note that the recent initial studies on computational prediction of post-translational folding pathways (mentioned in Section \ref{sect:introduction} and discussed in Section \ref{subsect:comp_pathway}) contain less-detailed 3D structural data capturing coarser-grained folding dynamics information for 30 proteins \cite{zhao2024foldpathreader,zhao2024recent}. Here, the only knowledge on folding dynamics is which portion of a protein's \emph{native} 3D structure folds first, and which folds second/last; information is available only for these two temporal states \cite{zhao2024foldpathreader,zhao2024recent}.
        
        \begin{figure*}
            \begin{center}
            \includegraphics[width=0.86\textwidth, trim= 0cm 16cm 0.25cm 0.5cm]{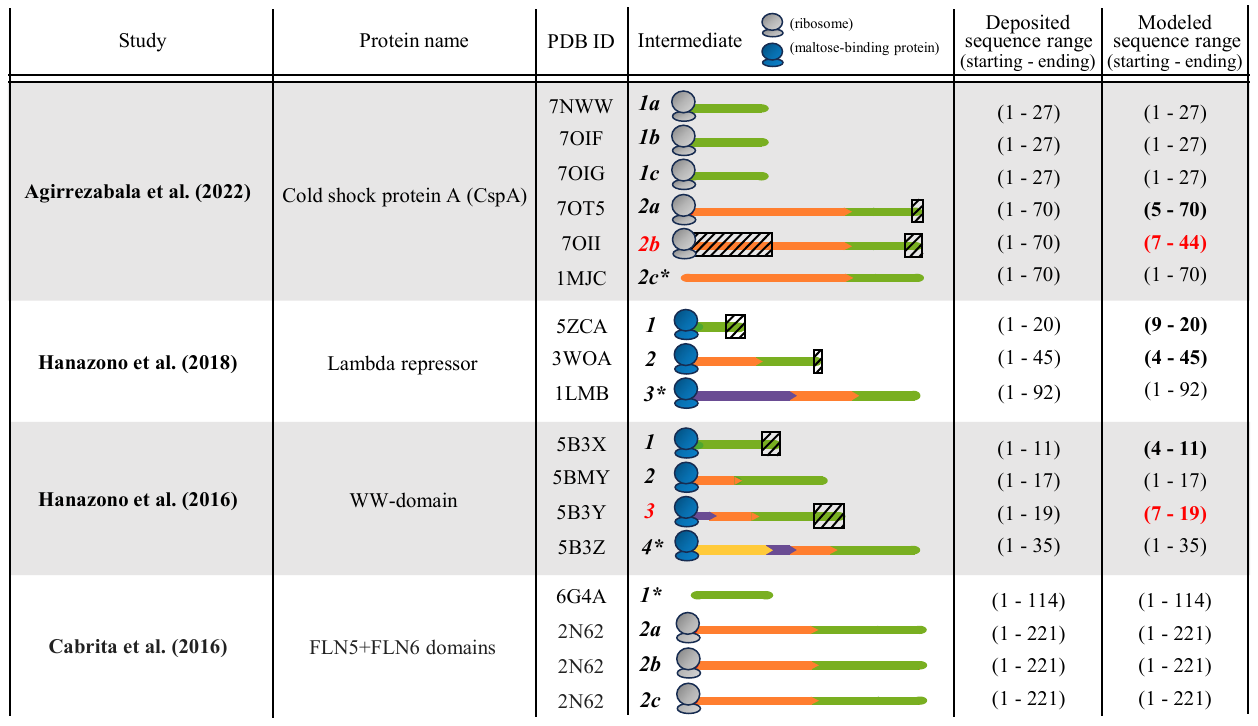}
            \end{center}
            \caption{Details of the four studies we have identified that report 3D structural data of co-translational intermediates. The protein from the first study listed has intermediates for two deposited sequence ranges (1: 1-27, 2: 1-70). For each of the two intermediates, three distinct conformations are provided (a, b, and c). Conformation 2c that is no longer  attached to the ribosome is the native structure of the protein (as denoted by an asterisk);  conformations 2a and 2b have the the same deposited sequence range as conformation 2c, but have structural differences to conformation 2c due to the appearance of the ribosome. Similar holds for the last study listed, whose protein has intermediates corresponding to two sequence ranges (1: 1-114, 2: 1-221). Intermediate 1 corresponds to a native structure of isolated (i.e., post-translationally folded, ribosome-independent) sequence region shown in green (corresponding to domain FLN5). Intermediate 2 corresponds to co-translationally folded sequence consisting of domains FLN5 (green) and FLN6 (orange) combined. For intermediate 2, three distinct confirmations are provided (a, b, and c; note that 3D structures of all three of these are available in the same PDB file, with the same PDB ID 2N62). For more information on why the shorter of the two intermediates is the native structure, see the last paragraph in Supplementary Section \ref{sect:sup_co_studies}. In the second and third studies listed, the given protein has three (1-3) and four (1-4) intermediates, respectively, of growing length, each with a single conformation. For both studies, the last intermediate is the native structure. The third column contains the PDB ID of each conformation of an intermediate. In the fourth column, each conformation is represented as a colored line where colors denote the same sequence region between intermediates. Also in the fourth column, the gray or blue entity at the C-terminus is a ribosome or maltose-binding protein, respectively; only the conformations with PDB IDs 1MJC and 6G4A do not have such an entity attached to them because these intermediates were experimentally studied in the absence of either entity. The fifth and sixth columns are defined in the same way as in Supplementary Fig. \ref{fig:pathway_data}. A modeled sequence range is bolded if a portion of the deposited sequence is missing from the modeled sequence; such a discrepant subsequence is represented with a hatched box overlaid on the deposited sequence in the fourth column (Supplementary Section \ref{sect:sup_tm_score} discusses how we handle 3D structural comparison in Section \ref{subsect:comp_cotrans} and later in the presence of minor subsequence discrepancies). A modeled sequence range is in red if its corresponding conformation is excluded from subsequent analyses in Section \ref{subsect:comp_cotrans} and later because of major subsequence discrepancies considered to affect data completeness/quality (described in Section \ref{subsect:exp_cotrans}).}
            %
            %
            \label{fig:cotrans_data}
        \end{figure*}

    \subsection{Experimental co-translational intermediates}\label{subsect:exp_cotrans}

        Our literature search on experimentally determined 3D structural data for co-translational intermediates identifies 
        only four studies that explicitly document 3D structures of intermediates in PDB: Agirrezabala et al. (2022) \cite{agirrezabala2022switch}, Hanazono et al. (2018) \cite{hanazono2018co}, Hanazono et al. (2016) \cite{hanazono2016structural},  and Cabrita et al. (2016) \cite{cabrita2016structural} (Fig. \ref{fig:cotrans_data} and Supplementary Section \ref{sect:sup_co_studies}). The current data on co-translational intermediates suffer from similar limitations as the data on post-translational intermediates -- each of the four studies on co-translational intermediates also reports a single protein, the studied proteins are also relatively short (all but one of the four studies analyze proteins with fewer than 100 residues), all four studies also provide quite few (2-4) intermediates. So, the current data on co-translational intermediates are also sparse. This further emphasizes the need for innovative experimental biophysical technologies for better capturing data on intermediates.
        
        Across the four studies/proteins, there are 11 distinct intermediates. Just as multiple 3D structural conformations may be reported for a native state, multiple conformations may be reported for an intermediate state as well. For three out of the 11 intermediates, three conformations have been reported per intermediate; for the remaining eight intermediates, only one conformation has been reported per intermediate  (Fig. \ref{fig:cotrans_data}). Hence, there are a total of 17 conformations for the 11 intermediates. For six of the conformations, their modeled sequences are shorter than their deposited sequences (PDB IDs 70T5, 70II, 5B3X, 5B3Y, 5ZCA, 3W0A). Of the six, we discard from further analyses the conformation of an intermediate with PDB ID 70II due to a huge data loss of $\sim$50\% in the modeled sequence compared to the deposited sequence. Also, we discard from further analyses the intermediate with PDB ID 5B3Y, because (i) this is the second largest data loss, (ii) this intermediate's deposited sequence adds only two extra amino acids (deposited sequence range of 1-19) to the intermediate with PDB ID 5BMY (deposited sequence range of 1-17), and (iii) 5B3Y loses many of its amino acids in the modeled sequence but 5BMY does not lose any; hence, 5BMY is an intermediate of much higher data quality, and again, almost as long as the deposited sequence of 5B3Y. We continue to analyze the remaining four conformations (i.e., 70T5, 5B3X, 5ZCA, 3W0A) with some minor data loss. In total, we proceed with analyzing 15 conformations for 10 intermediates (i.e. all but the two red ones in Fig. \ref{fig:cotrans_data}).

        We measure  structural similarities (TM-scores) between all pairs of (partial and full) conformations of intermediates of the same protein (Supplementary Section \ref{sect:sup_tm_score}). We find the following (Supplementary Fig. \ref{fig:co_trans_n_TMscore_mtx}). 
        First, we focus on distinct conformations for the same intermediate (e.g., 2a vs. 2b vs. 2c from a given study). All corresponding TM-scores are below 0.5, indicating that no two conformations of the same intermediate have the same fold. Second, we focus on the conformational change of a sequence of an intermediate gradually over time (e.g., a green sequence in an intermediate vs. the same green sequence in the next intermediate for the same protein). Overall, we observe quite a large conformational change of the same sequence between consecutive time points in the presence of additional amino acids being translated and added to the 3D structure, often resulting in a changed fold (i.e. TM-score below 0.5) of the given sequence during translation. Third, to evaluate the effect of time passed, we compare conformations corresponding to the same sequence of an intermediate at closer vs. more distant times (e.g., a sequence from time 1 to time 2 vs. the same sequence from time 1 to time 3). In all comparisons, more time-distant conformations of intermediates tend to show higher levels of conformational change than time-closer conformations for the same protein. For more detailed results, see Supplementary Section \ref{sect:sup_exp_cotrans_deates_2}. 

        For completeness, we summarize the aims of the four studies from this section in more detail. Due to space constraints, we report this in Supplementary  Section \ref{sect:sup_co_studies}. 

\section{Computational 3D structural data on protein folding pathway intermediates}\label{sect:comp_data}

    \subsection{Computational post-translational intermediates}\label{subsect:comp_pathway}

        With advancements in protein 3D structure prediction, available data on native structures has increased drastically. Prior to the release of the AlphaFold database, $\sim$200,000 experimentally determined 3D structures were reported in PDB -- just a fraction of the billions of known protein sequences \cite{varadi2024alphafold}. Now, computationally predicted 3D structures are available for over 214 million proteins \cite{varadi2024alphafold}. Given the lack of experimentally determined 3D structural data on folding intermediates, can existing methods for predicting native 3D structures accurately predict 3D structures of intermediates? 
           
        Outeiral et al. (2022) investigated this question for eight prominent existing methods, including AlphaFold2, on \textit{post-translational} intermediates \cite{Outeiral2022CurrentSP}. 
        Each of 170 analyzed proteins  was assigned one of two folding kinetics classes: \cite{manavalan2019pfdb,pancsa2016start2fold} two-state (90/170) or multi-state (80/170). 79 of the 90 two-state proteins had additional kinetics data -- folding rate constants. To predict a protein's folding pathway, the existing methods were used to output intermediate 3D structures produced during the process of predicting the native structure \cite{Outeiral2022CurrentSP}. The seven methods (except AlphaFold2) were used to predict 10 pathways per protein. Five of the more scalable methods were also used to predict 200 pathways per protein. For AlphaFold2, only one predicted pathway was available per protein. We could not find information on the number of intermediates per pathway, nor the predicted structures of the intermediates, reported in the paper by Outeiral et al.  (2022).
        
        When evaluating a method's predicted pathway, Outeiral et al. (2022) simply assumed that many proteins fold by first forming secondary structures, followed by forming tertiary contacts between them. So, a predicted pathway was examined using its predicted 3D structural intermediates to track when native contacts formed between each pair of secondary structure elements over time. If all pairs formed their native contacts around the same time, this indicated a two-state folding mechanism, where folding occurs in a single concerted step. In contrast, if some pairs formed contacts earlier and others later, this suggested a multi-state folding mechanism. This information was then used to evaluate whether a method's predicted pathways (i) are predictive of a protein's folding kinetics class (i.e. two-state or multi-state), and (ii) correlate with experimentally measured folding rate constants.
        
        For the first evaluation task, all structure prediction methods achieved statistically significant yet quite modest performance (with the area under the receiver-operating characteristic curve, i.e., AUROC, scores between 0.56 and 0.675, when predicting 10 pathways per method per protein). The methods were compared to a trivial baseline relying only on chain length; chain length served as a deliberately simple, sequence-agnostic baseline to test whether structure prediction methods actually learned meaningful folding physics. This simple baseline outperformed all structure prediction methods (AUROC score of 0.739). Results in the first evaluation task were qualitatively similar for the other performance accuracy measures and when predicting 200 pathways per method per protein (Supplementary Section \ref{sect:sup_comp_posttrans_deates} and Supplementary Table \ref{tab:outerial_study_results}). For the second evaluation task, chain length had the strongest and correctly-signed correlation, again outperforming any structure prediction method (Supplementary  Section \ref{sect:sup_comp_posttrans_deates} and Supplementary Fig. \ref{fig:outerial_study_results_2}).
        
        These findings suggested that folding pathways predicted by the existing structure prediction methods fail to meaningfully capture folding kinetics data or correlate with experimental folding rate data. In other words, the existing methods fail to accurately model post-translational folding pathways.

        The existing methods, originally designed to predict native 3D structures, were  applied by Outeiral et al. (2022) \cite{Outeiral2022CurrentSP} in a naive way to generate pathways that yielded the native structure, without any direct capability to capture protein folding dynamics. Instead,  the recent  Pathfinder \cite{Huang2023PathfinderPF} is a pioneering method explicitly designed to predict a post-translational folding pathway, i.e., to capture the dynamics of post-translational folding; it is a Monte Carlo  approach that uses conformational sampling to explore the transition probabilities of folding intermediates and infer likely folding pathways. To assess whether a protein's post-translational pathway predicted by Pathfinder captured folding dynamics well, the contact order (average sequence distance between residues that form native contacts) was computed in each intermediate of the pathway. Then, by \textit{qualitatively} analyzing how contact order evolved over the course of the predicted pathway and comparing this information to experimentally observed folding data, it was evaluated whether the predicted pathway ``sufficiently'' captured realistic folding dynamics (``sufficiently'' because the evaluation was qualitative/subjective, based on descriptive folding information from the literature, rather than quantitative/objective). Pathfinder predicted folding pathways ``sufficiently'' well for 11 proteins, while it did not do well for six other proteins \cite{Huang2023PathfinderPF}.
        
        In parallel to Pathfinder, the same lab proposed PAthreader \cite{zhao2023protein}, which is based on remote homologous template recognition. Their more recent approach, FoldPAthreader \cite{zhao2024foldpathreader}, combines the  ideas of Pathfinder and PAthreader. When evaluated on the 30 proteins with only coarse-grained folding dynamics information on which portion of the \emph{native} 3D structure folds first vs. second/last (Section \ref{subsect:exp_pathway}), for 21 of them, FoldPAthreader's predicted post-translational pathways were judged as being consistent with that information.

        Monte Carlo approaches such as Pathfinder complement other common computational strategies such as molecular dynamics simulations (which provide atomistic detail through force field-driven trajectories) and coarse-grained models (which simplify molecular representations to enable broader exploration of folding behavior) \cite{kmiecik2016coarse}. Despite their differences, all three approach types share limitations, specifically a reliance on highly parameterized energy functions, simplified assumptions (e.g. modeling only backbone or only side-chain atoms, or grouping several atoms into a single unit), or constrained timescales, which can introduce significant biases in predictions of protein folding dynamics and a limited simulated timescale \cite{Gershenson2019SuccessesAC, hu2017critical, kmiecik2016coarse}. Moreover, the existing approaches, including those from the Outeiral et al. (2022) \cite{Outeiral2022CurrentSP} and Pathfinder \cite{Huang2023PathfinderPF} studies, rely on experimental observations of secondary structure formation or folding kinetics information for validation -- rather than direct comparison to 3D structures of intermediates (as such data do not exist for post-translational intermediates other than the two studies mentioned in Section \ref{subsect:exp_pathway}). Same holds for the FoldPAthreader approach. And while its evaluation did get somewhat less qualitative/subjective and more quantitative/objective, still only coarse-grained 3D structural data capturing limited folding dynamics information on top of the native structures was used \cite{zhao2024foldpathreader}. We imagine that having actual 3D structures of post-translational intermediates along a folding pathway could help better evaluate and refine post-translational folding pathway prediction tools.

    \subsection{Computational co-translational intermediates}\label{subsect:comp_cotrans}

        \subsubsection{Can AlphaFold2, an established method for predicting native 3D structures, predict  3D structures of co-translational intermediates?}\label{subsubsect:predict_cotrans}

            Despite significant advances in protein structure prediction, our literature search has uncovered no computational methods for predicting 3D structures of co-translational intermediates -- with the exception of two studies \cite{wang2025cotranslational,tao2025pathway} published after submission of this paper of ours. So, as an original contribution to this paper, we explore whether AlphaFold2, a prominent existing method designed for prediction of native 3D structures, can predict 3D structures of co-translational intermediates -- following a similar question that Outeiral et al. (2022) \cite{Outeiral2022CurrentSP} investigated for post-translational intermediates. While the studies from Section \ref{subsect:comp_pathway} evaluated predicted post-translational intermediates with respect to quantitative measures (e.g. folding kinetics data) or qualitative information (e.g. secondary-structure formation information), we cannot do the same for predicted co-translational intermediates, because such data are limited for co-translational intermediates (Section \ref{sect:introduction}). Instead, we predict 3D structures of co-translational intermediates for the four proteins that have experimentally determined 3D structures of co-translational intermediates available (Section \ref{subsect:exp_cotrans}). Then, we evaluate the predictions by directly comparing them to the 3D structures of experimentally determined co-translational intermediates. In this analysis, we rely on the  15 considered conformations for 10 intermediates (i.e. all but the two red ones in Fig. \ref{fig:cotrans_data}), which we have already cleaned to remove data noise (Section \ref{subsect:exp_cotrans}), hence hopefully yielding higher-confidence results from our analyses. We predict 3D structures of the co-translational intermediates as follows.
            
            \begin{figure*}
                \begin{minipage}[c]{0.57\textwidth}
                    \includegraphics[width=\textwidth,trim= 0cm 20.4cm 8cm 0.22cm]{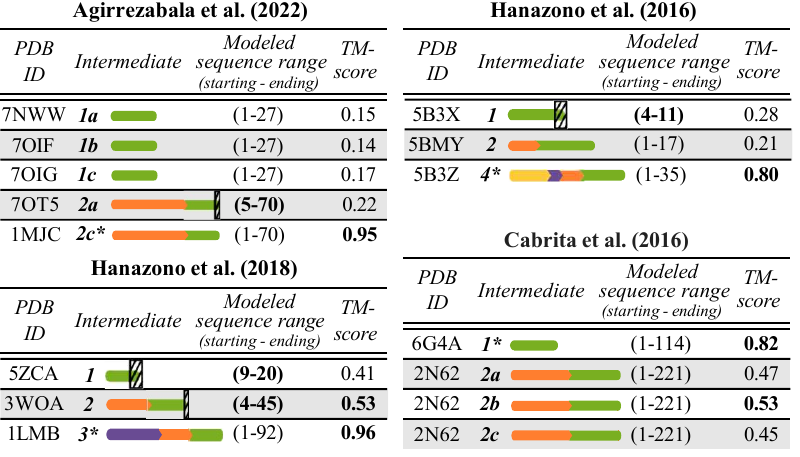}
                \end{minipage}\hfill
                \begin{minipage}[c]{0.43\textwidth}
                    \caption{Structural similarities in terms of TM-scores between the AlphaFold2-predicted vs. experimentally determined structures for the 15 considered conformations of the 10 co-translational intermediates (i.e. for all but the two red ones in Fig. \ref{fig:cotrans_data}). Each of the four studies, has a corresponding table; the first three table columns are already explained in Fig. \ref{fig:cotrans_data} (except that here we do not show or analyze an attached entity at the C-terminus). The fourth table column reports the TM-score between an AlphaFold2's prediction and the corresponding experimentally determined confirmation; all TM-scores above 0.50 are bolded, indicating the same overall fold. 3D visualizations of the AlphaFold2-predicted structures are shown in Supplementary Fig. \ref{fig:supp_AF2_n_TMscore}. Note that each result here is for the highest-ranked AlphaFold2-predicted structure; results for all five highest-ranked AlphaFold2-predicted structures are shown in Supplementary Fig. \ref{fig:supp_af2_tm_score_rank2-5}.}
                    \label{fig:AF2_n_TMscore}
                    %
                    %
                \end{minipage}
            \end{figure*}
            
\begin{figure*}[ht]
        \begin{center}
                \includegraphics[
                    width=0.95\textwidth,
                    trim= 0.5cm 10cm 3.6cm 0cm,
                    clip
                ]{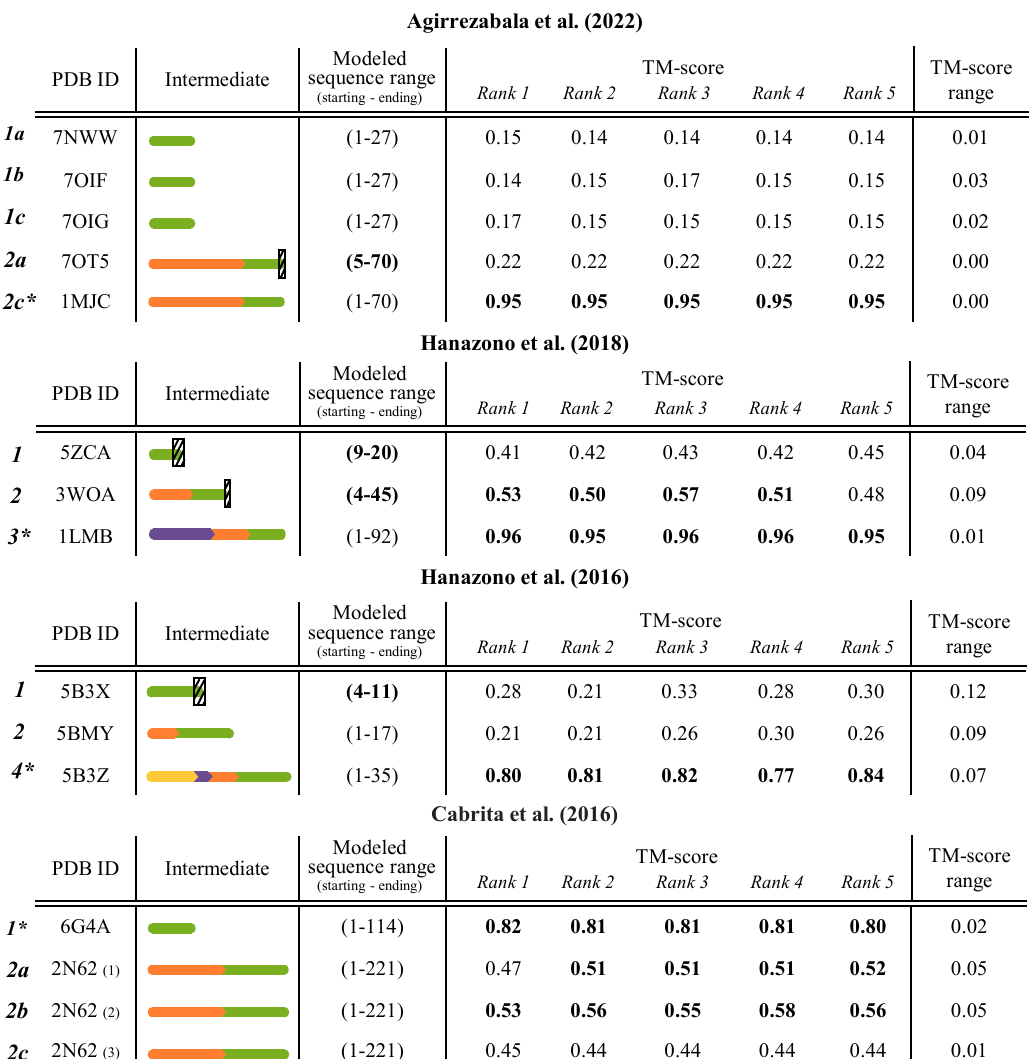}%
        \end{center}
        \caption{The same type of results as in Fig. \ref{fig:AF2_n_TMscore} in the main paper, except that we report TM-scores for all five top-ranked AlphaFold2-predicted structures. Specifically, the figure shows structural similarities in terms of TM-scores between Alphafold2-predicted structures and experimentally determined structures, for the  15 considered conformations of the 10 co-translational intermediates (i.e. for all but the two red ones in Fig. \ref{fig:cotrans_data} in the main paper). For each of the four studies, there is a corresponding table; the first three table columns are already explained in Fig. \ref{fig:cotrans_data} in the main paper (except that here we do not show or analyze an attached entity at the C-terminus). The fourth table column reports the TM-scores of all five highest-ranked predicted structures from AlphaFold2. All TM-scores with values higher than  $0.50$ are bolded, corresponding to structures that have the same overall fold. The fifth (last) column reports the maximum absolute TM-score difference between all pairs of the five AlphaFold2-predicted structures for a given (conformation of the) intermediate. }
        \label{fig:supp_af2_tm_score_rank2-5}
    \end{figure*}

            For each of the 15 conformations of the intermediates, we give its deposited sequence (excluding the ribosome or maltose-binding protein) as input into AlphaFold2 (ColabFold \cite{mirdita2022colabfold} v1.5.2). As a result, each conformation has a corresponding AlphaFold2-predicted 3D structure. AlphaFold2 generates five predicted structures for each sequence, ranked by confidence. We report results for the top-ranked predicted structure. Nonetheless, results for all five AlphaFold2-predicted structures of a given sequence are  qualitatively similar and quantitatively almost identical regardless of which AlphaFold2-predicted structure for a given sequence is used (Supplementary Fig. \ref{fig:supp_af2_tm_score_rank2-5}). Note that we use AlphaFold2 rather than AlphaFold3 because the main improvements of the latter compared to the former focus on modeling protein complexes and biomolecular interactions, rather than offering tremendously significant advances for predicting the native structure of a single-chain protein \cite{abramson2024accurate, elofsson2025alphafold3}. 

            To evaluate how well AlphaFold2 predicts 3D structures of co-translational intermediates, for each conformation, we compare its experimentally determined structure to its AlphaFold2-predicted structure using TM-score (Supplementary Section \ref{sect:sup_tm_score}). Our findings are as follows (Fig. \ref{fig:AF2_n_TMscore}). Unsurprisingly, AlphaFold2 performs well for native conformations of intermediates (TM-scores in the 0.80-0.96 range, depending on the protein). There is a stark contrast in its performance for non-native intermediates, where it does not work as well. Namely, of the 15 conformations of the intermediates that are evaluated, 11 are non-native. For nine of these, AlphaFold2 yields TM-scores below $0.5$, specifically in the 0.14-0.47 range (indicating different folds), and for the remaining two non-native conformations, AlphaFold2 yields TM-scores of just above 0.5. Due to space constraints, for key observations per study, see Supplementary Section \ref{sect:sup_comp_posttrans_deates_4}.
            
            Our findings reinforce that current computational structure prediction methods (like AlphaFold2) that are designed to predict native 3D structures are not inherently able to capture well co-translational intermediates. This observation is consistent with the previous findings for post-translational intermediates \cite{Outeiral2022CurrentSP} (Section \ref{subsect:comp_pathway}).
            
            \begin{figure*}
                \begin{center}
                \includegraphics[width=\textwidth,trim= 0cm 14.85cm 0cm 2.25cm]{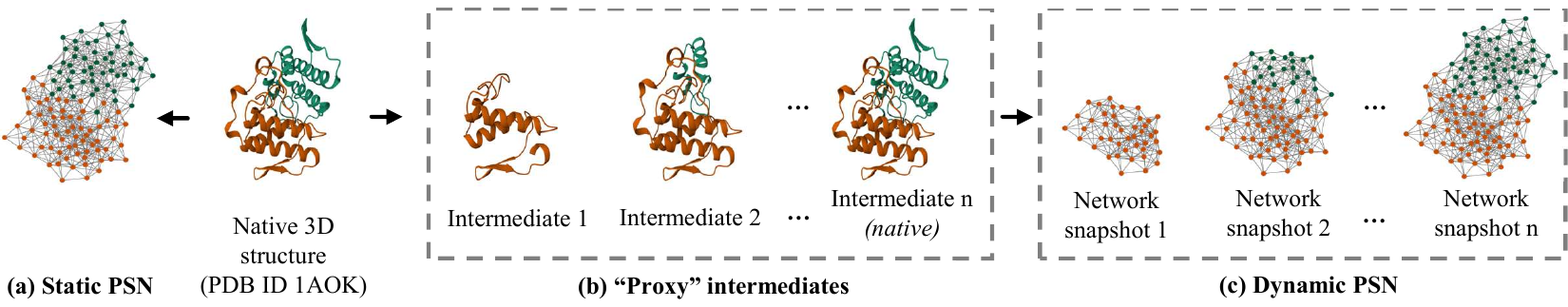}
                \end{center}
                    \caption{Illustration of \textbf{(a)} a static protein structure network (PSN), \textbf{(b)} ``proxy'' co-translational intermediates, and \textbf{(c)} dynamic PSN, all for the same protein with PDB ID 1AOK. The static PSN captures the native structure of the protein. ``Proxy'' co-translational intermediates capture gradually increasing portion of the protein sequence and the corresponding portion of the native 3D structure, with the last ``proxy'' intermediate capturing the entire sequence and native structure. Each ``proxy'' intermediate (intermediate 1, 2, ..., n) corresponds to a network snapshot (network snapshot 1, 2, ..., n, respectively) in the dynamic PSN. The figure has been adapted from Newaz et al. (2022) \cite{newaz2022multi}.}
                %
                %
                \label{fig:psn_types}
            \end{figure*}
        \subsubsection{Inferring ``proxy''  co-translational intermediates}\label{subsubsect:co-trans_proxy}
     
            Computational analyses of \emph{post-translational} folding have been done on at least somewhat large scale for dozens to hundreds of proteins -- though on non-3D structural data (Section \ref{subsect:comp_pathway}). For \emph{co-translational} folding, we could only analyze four proteins for which 3D structures of their intermediates are available, since no quantitative nor qualitative data exist on co-translational intermediates (Section \ref{subsubsect:predict_cotrans}). This raises the question of whether 3D structures of co-translational intermediates can be predicted on a larger scale, and if so, how they would be validated. Our analysis in Section \ref{subsubsect:predict_cotrans} reveals that one of the best existing methods designed to predict native 3D structures --  AlphaFold2 -- is not equipped to capture structures of co-translational intermediates. Certainly, testing other method types \cite{Huang2023PathfinderPF,zhao2024foldpathreader,wang2025cotranslational,tao2025pathway} in the task of predicting 3D structures of co-translational intermediates could be a valuable future direction, although validation would be challenging given the lack of data on co-translational intermediates. So, how may it be possible to predict and validate co-translational intermediates on a much larger scale? In our recent work, we took a step forward in this direction, by utilizing a bulk of data that is readily available -- native 3D structures -- to infer ``proxy'' co-translational intermediates on a large scale \cite{newaz2022multi}.

            Briefly, we extracted a protein's 3D structural ``proxy'' co-translational intermediates from its native structure as follows \cite{newaz2022multi}. Motivated by the co-translation process (Fig. \ref{fig:pathway_n_cotrans}b), the protein's first ``proxy'' intermediate contains the first $k$ amino acids of the entire sequence (and the corresponding substructure of the native structure), the second ``proxy'' intermediate  contains the first $2k$ amino acids of the entire sequence (and the corresponding substructure of the native structure), etc; this continues until arriving at the last ``proxy'' intermediate, which captures the entire protein's sequence (and the entire native 3D structure) \cite{newaz2022multi}. For discussion of the best choice of $k$, see Supplementary Fig. \ref{fig:supp_k_val_alt}. The set of all ``proxy'' intermediates for a protein aims to mimic the process of co-translation. We say ``mimic'', because unfortunately this procedure is unable to capture any conformational changes of an intermediate over time, i.e. any actual dynamics of the co-translational folding process, because all of the intermediates are extracted from the same, static native structure. Although these intermediates do not explicitly model co-translational folding as it occurs in vivo (hence the term ``proxy''), they were the most practical large-scale method for mimicking co-translational folding, i.e. the best one could do on a large scale at the moment without significant computational and methodological innovation.
            
            We evaluated our approach for extracting ``proxy'' structural intermediates in the task of protein structure classification (PSC) \cite{koehl2006protein,newaz2020network}, which is closely related to protein function prediction \cite{gligorijevic2021structure}. Traditional 3D-structural features for PSC are extracted directly from 3D structures. Instead, one can first model a 3D structure as a protein structure network (PSN) \cite{faisal2017grafene}. Then, a wealth of network-based methods can be used to extract network features for use in the PSC task (which otherwise could not be extracted directly from 3D protein structures). Our lab had already demonstrated that using PSN features extracted from native structures often proved to be more accurate and typically faster than using sequence and non-network-based 3D structural approaches in the PSC task \cite{newaz2020network}. However, traditional PSN approaches (including our lab's previous approaches)  had modeled a native 3D structure as a \emph{static} PSN  (Fig. \ref{fig:psn_types}a), because experimental data on dynamics (i.e. on co-translational intermediates) are lacking (Section \ref{subsect:exp_cotrans}). However, with the availability of ``proxy'' co-translational intermediates, more recently, we were able to model a native 3D structure as a \emph{dynamic} (Fig. \ref{fig:psn_types}c) rather than static PSN \cite{newaz2022multi}; the desire to model a native 3D structure as a dynamic PSN, in order to capture, even implicitly, the dynamics of protein folding, was the reason why we came up with the idea of ``proxy'' co-translational intermediates. We demonstrated that modeling a 3D structure as a dynamic PSN \cite{newaz2022multi, newaz2019graphlets} performed better in the PSC task than modeling it as a static PSN. These results confirmed that considering dynamic information, even only as ``proxy'' co-translational intermediates derived from native structures, could improve insights into folding-related phenomena.

         \subsubsection{``Proxy'' vs. AlphaFold2-predicted co-translational intermediates}\label{subsubsect:proxy_AF2}

            \begin{table*}[h]
                \begin{minipage}[c]{0.61\textwidth}
                \includegraphics[width=\textwidth,trim= 0cm 23.5cm 7.5cm 0.5cm]{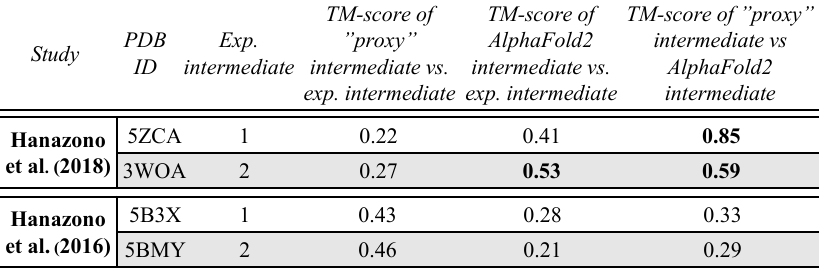}
                \end{minipage}\hfill
                \begin{minipage}[c]{0.38\textwidth}
                \caption{Structural similarity (with respect to TM-score)  of  experimentally  determined (exp.) vs. ``proxy'' vs.  AlphaFold2-predicted co-translational intermediates. For reasons stated in the text, this analysis is done on two of the four considered studies/proteins:  Hanazono et al. (2018) \cite{hanazono2018co} and Hanazono et al. (2016) \cite{hanazono2016structural}. For each study/protein, results are shown for all (both) of its \emph{non-native} intermediates. All TM-scores with values higher than 0.50 are bolded, corresponding to structures that have the same overall fold.
                }
                \label{tab:supp_af2_proxy_compare}
                \end{minipage}
            \end{table*}

            The PSC task is not explicitly designed to address protein folding \emph{dynamics}. So, as an original contribution to this current study, we evaluate ``proxy'' intermediates in a different task more directly related to folding dynamics. Specifically, we assess 3D structural similarity (as measured by TM-score) between  experimentally determined intermediates and their corresponding ``proxy'' intermediates. We do not anticipate a perfect structural match between the two. This is because the former often undergo conformational changes during co-translation (as the TM-scores from Supplementary Fig. \ref{fig:co_trans_n_TMscore_mtx} show), but this is not reflected in the design of the latter (as the ``proxy'' intermediates are extracted from a native structure). So, the comparison between experimentally determined intermediates and ``proxy'' intermediates is not our key goal here. Instead, it is to evaluate whether the experimentally determined intermediate conformations are structurally matched better or worse by their corresponding ``proxy'' intermediates or by their corresponding AlphaFold2-predicted intermediates. If better, this would further highlight AlphaFold2's limitations in modeling protein folding dynamics.

            When we analyze a subset of the four studies/proteins with available experimentally determined 3D structures of co-translational intermediates that meet relevant criteria (Supplementary Section \ref{sect:sup_comp_posttrans_deates_5}), our findings are as follows (Table \ref{tab:supp_af2_proxy_compare}). The AlphaFold2-predicted intermediates match the experimentally derived intermediates just as poorly as the ``proxy'' intermediates.
            All TM-scores are quite low, between $0.21-0.53$. AlphaFold2 performs better (yet not too well, as indicated by the TM-scores) for a half of the analyzed data, while ``proxy'' intermediates perform better (also not too well) for the other half. An additional observation is that depending on the protein, AlphaFold2's predictions for non-native intermediates are of the same fold as proxy intermediates (TM-scores in the 0.59-0.85 range) in some cases, but of different folds than proxy intermediates (TM-scores in the 0.28-0.33 range) in other cases (Table \ref{tab:supp_af2_proxy_compare}).

            We showed in our recent study \cite{newaz2022multi} that using ``proxy'' intermediates in a dynamic fashion is a powerful method for capturing more information from native 3D structures in the task of PSC than analyzing the native structures in a traditional, static fashion (Section \ref{subsubsect:co-trans_proxy}). In the above analysis, we have shown that ``proxy'' intermediates perform comparably to AlphaFold2 in the task more closely related to protein folding dynamics -- predicting 3D structures of co-translational intermediates. Again, we do not aim to imply that ``proxy'' intermediates are an appropriate approach for modeling 3D structures of folding intermediates, and thus for capturing  folding dynamics. Instead, our findings imply that AlphaFold2 does not consistently outperform the baseline ``proxy'' intermediates in the second task, and it also most often yields TM-scores below 0.5 that indicate a changed fold, all of which underscores the need for new types of approaches that will be capable of accurately modeling folding dynamics and predicting 3D structures of intermediates. 

\section{Discussion and concluding remarks}\label{sect:discussion}

Our study (and others) highlights a fundamental barrier to advancing the understanding of protein folding dynamics: the near-total absence of 3D structural data on folding intermediates, in the context of both post-translational and co-translational folding. For post-translational folding, our literature search has revealed only two studies that report experimentally determined 3D structures of post-translational intermediates (we have also identified 30 proteins with less-detailed 3D structural data capturing coarser-grained folding dynamics information on which portion of a protein's \emph{native} 3D structure folds first vs. second/last \cite{zhao2024recent,zhao2024foldpathreader}). For co-translational folding, we have identified  only four studies that report experimentally determined 3D structures of co-translational intermediates. Each of these two plus four proteins is constrained by a limited number of intermediates, and most by short length. 

While thermodynamic and kinetic datasets do exist on a somewhat broader scale for post-translational folding, they typically provide a single quantity per protein rather than per intermediate along a folding pathway. Even in the best cases, where some proteins have a measurement per intermediate, the data are only quantitative, lacking 3D spatial information. For co-translational folding, the situation is even more sparse; there are no centralized resources containing data of these types; instead, only a few isolated studies provide such data, typically with one protein per study. As a result, while thermodynamic and kinetic data allow for quantitative analyses of post-translational folding, their lack of accompanying 3D structural information constrains the depth of such analyses. For co-translational folding, the near-complete absence of both quantitative and 3D structural data for intermediates leaves little means for any analysis.

Given the lack of experimentally determined 3D structural data on intermediates, it is natural to ask whether computational methods -- especially those that have demonstrated success in predicting native 3D structures -- can help bridge the gap. Our findings from Section \ref{subsect:comp_cotrans}, along with those of Outeiral et al. (2022) \cite{Outeiral2022CurrentSP} discussed in Section \ref{subsect:comp_pathway}, provide a strong caution against this assumption. These findings reinforce the conclusion that existing structural prediction methods do not generalize well outside of native conformations, and thus cannot be expected to accurately predict 3D structures of intermediates along a folding pathway.

So, how may this be achieved (more) accurately? One possibility may be to retrain existing structure prediction methods with data on folding intermediates, and in doing so, re-engineer these approaches to predict the 3D structures of folding intermediates. However, this direction is limited by the near-total lack of experimentally determined 3D structural data on intermediates, both for post-translational and co-translational folding. 

Another possibility may be running AlphaFold2 using advanced input adjustments to generate an ensemble of structural models for full-length sequences \cite{wallner2023afsample,raouraoua2024massivefold,monteiro2024high}. While approaches like these can potentially reveal multiple plausible conformations, we hypothesize that they are unlikely to be capable of accurately generating non-native intermediate structures (for either a partially folded or full-length sequence); fundamentally, these models are trained to predict stable, native structures based on multiple sequence alignment information. Moreover, beyond these advanced ways to run AlphaFold2 (or any other existing method for predicting native 3D structures), there also exist specialized AI models explicitly designed to generate diverse protein conformations in the task of protein conformational ensemble prediction \cite{jing2023eigenfold,wang2024protein,lewis2025scalable,jing2024alphafold}. However, because unstable non-native intermediate structures can differ significantly from native conformations, we hypothesize that these tools are also likely ill-suited for predicting protein folding pathways. Although we believe both of these hypotheses to be true,  definitive verification would require empirical testing; this is outside the scope of the current study and is a subject of future work.

Yet another possibility may be to develop methods that more accurately account for the underlying biochemical and physical factors driving protein folding dynamics, such as the geometry of the ribosome (and its tunnel) or other mechanisms by which the ribosome thermodynamically regulates co-translational folding, presence of chaperones, or vectorial, time-dependent nature of translation (in the case of co-translational folding) \cite{agirrezabala2022switch, chan2022ribosome, nilsson2015cotranslational, wang2025cotranslational}. In a bit more detail regarding the importance of the ribosome’s role: it is understood that disruptions to translation by the ribosome -- such as artificially induced pauses -- can trigger widespread protein misfolding and aggregation, underscoring how tightly coordinated elongation and folding must be. The ribosome helps maintain this coordination by preventing premature co-translational misfolding by the unfolded nascent chain, lowering the entropic penalty of protein folding, and participating with chaperones in assisting folding even before protein synthesis is complete \cite{duran2025native}. Complementing these biochemical observations, simulation studies that explicitly model the ribosomal exit tunnel and stepwise translation have predicted how the ribosome influences the conformations adopted by nascent peptides at the moment of release -- structures that typically differ  from their native fold \cite{tao2025pathway}.

In addition to the above biochemical and
physical factors driving protein folding dynamics, incorporating codon usage patterns might improve the prediction of co-translational folding pathways because it serves as a proxy for translation rates, which are evolutionarily tuned to regulate the timing of protein synthesis \cite{o2014understanding}. For instance, slower translation rates allow the N-terminal portion of a protein to fold independently before the synthesis of the C-terminal portion is complete, while faster translation rates can help prevent the buildup of certain folding intermediates \cite{moss2024effects}. Even synonymous codon usage -- where different codons code for the same amino acid -- can affect translation speed and, as a result, impact co-translational folding \cite{moss2024effects,newaz2020codon}.

Recent efforts have begun to demonstrate the potential of considering these factors for better prediction of folding pathways. For post-translational folding, Pathfinder \cite{Huang2023PathfinderPF}, PAthreader \cite{zhao2023protein}, and FoldPAthreader \cite{zhao2024foldpathreader} were introduced (Section \ref{subsect:comp_pathway}).
For co-translational folding, we and others have modeled the vectorial nature of protein synthesis by explicitly capturing increasingly longer subsequences of a protein and then predicting their 3D substructures as the structures of co-translational intermediates; the latter has been achieved in four distinct ways, as follows. 

First,  we previously introduced ``proxy'' intermediates  \cite{newaz2022multi} -- 3D substructures representing progressively longer C-terminal sequence fragments of a protein extracted \emph{naively from its native structure} -- to model co-translational folding; while  ``proxy'' intermediates do not  capture protein folding dynamics \emph{explicitly}, we showed that their analysis is nonetheless significantly more accurate than traditional analyses of only the native structures in the task of predicting protein structural classes (Section \ref{subsubsect:co-trans_proxy}). 

Second, in the current study, we have used a more advanced idea of predicting 3D structures of co-translational intermediates by feeding progressively longer C-terminal sequence fragments of a protein into AlphaFold2; however, this approach was unable to accurately predict the existing experimentally determined co-translational intermediates (Section \ref{subsubsect:predict_cotrans}), likely because it does not effectively learn the kinetics (and thus dynamics) of protein folding \cite{Outeiral2022CurrentSP}. 

Third, a different existing structure prediction method  -- which predicts protein folding pathways while accounting for \emph{non-native} interactions and which considers some key principles of co-translational folding -- was used on increasingly longer subsequences of a protein to predict their 3D structures. Namely, the study by Wang et al. (2025) \cite{wang2025cotranslational} -- which we again note appeared as \emph{after} the completion of all of our analyses -- used Monte Carlo simulations that ``recapitulate intrinsic properties of co-translational folding, such as the sequential emergence of the peptide and its propensity to form secondary and tertiary structures'' and also ``provide insights into the thermodynamic stability and dimensions of folding intermediates, which determine whether structural elements can form inside the narrow ribosome exit tunnel'' \cite{wang2025cotranslational}. To evaluate the predicted 3D structures resulting from these simulations, Wang et al. (2025) performed in vitro experiments using truncated protein domains (sequence fragments of increasing length) attached to a ribosome that mimicked different stages of co-translation. The experimentally measured structural compactness and stability of these domains were then compared to the corresponding simulation-predicted conformations and stability profiles, and the two seemed to align. 

Fourth, another simulation-based approach that explicitly models the differences between co-translational folding and post-translational (free) folding is the framework introduced by Tao et al. (2025) \cite{tao2025pathway}. This study -- which also appeared \emph{after} the submission of our paper -- developed a framework that incorporates both a simplified geometric model of the ribosomal exit tunnel and a residue-by-residue translation process, enabling atomistic molecular dynamics  simulation of both co-translational folding and free folding of a protein's 3D structure. This approach was designed to address two specific questions. One question was what structure the nascent peptide adopts upon expulsion from the ribosomal exit tunnel. Using over eight milliseconds of simulations across three proteins with distinct topologies, Tao et al. (2025) found that co-translational folding influenced the nascent chain to adopt a more helix-rich structure with less long-range contacts upon exit from the ribosome compared to post-translational folding. The other question was how this structure  evolves during subsequent folding. By comparing co-translational and post-translational trajectories, subsequent folding pathways were found to be very similar, but with co-translational folding favoring conformations that promote faster folding. Note that Tao et al. (2025) did not evaluate/validate their simulated co-translational pathways against experimentally determined 3D intermediate structures; rather, they used their simulated pathways to examine how ribosomal confinement and protein translation bias early conformations and pathway selection. By accounting for these biases, in vivo observations that once appeared inconsistent with in vitro observations -- such as differences in folding efficiency or misfolding propensities -- could be understood as natural consequences of ribosome-guided pathway selection.

When a new method becomes available for predicting the 3D structures of folding intermediates, an immediate challenge will be the evaluation of its predictions, i.e., assessment of whether the predicted 3D structural intermediates match biological reality. This can be approached in several ways.

\begin{figure*}
        \begin{center}
        \includegraphics[width=\textwidth,trim= 0cm 9.55cm 4.25cm 0.25cm]{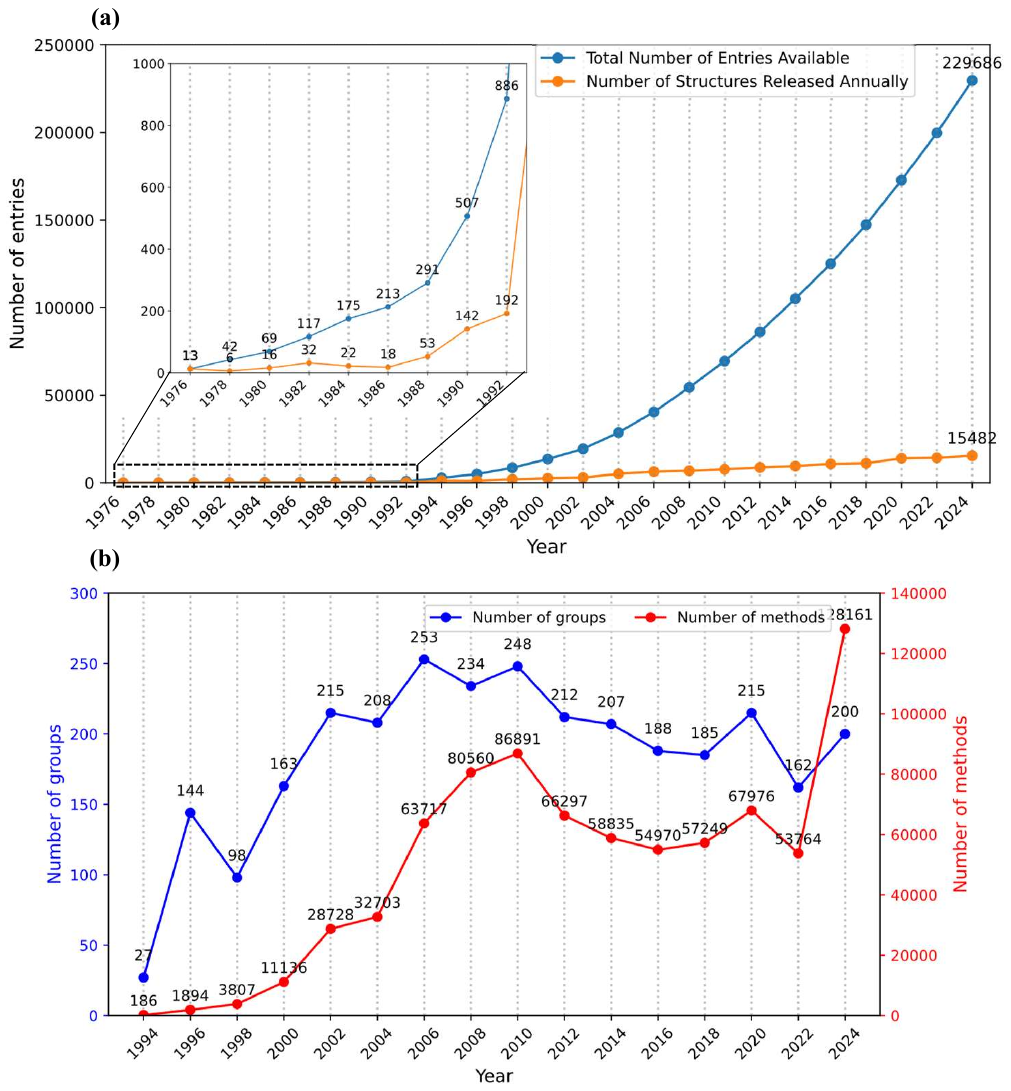}
        \end{center}
        \caption{The growth of \textbf{(a)} available protein 3D structures (i.e. entries) in PDB over time and \textbf{(b)} participating research groups and submitted computational predictive methods (also referred to as ``models'' in some literature) in the CASP competition over time. We collected these statistics from \url{https://www.rcsb.org/stats} and \url{https://predictioncenter.org}, respectively.}
        \label{fig:supp_pdb_n_casp}
    \end{figure*}

First, ideally, evaluation would involve 3D structural comparison between predicted and experimentally determined folding pathways. Although such experimental data are scarce, they are not entirely absent; such data exist for post-translational as well as co-translational folding for a handful of proteins (Section \ref{sect:exp_data}); this is similar to the initial scarcity of available protein 3D structures at the inception of PDB (see below and Supplementary Fig. \ref{fig:supp_pdb_n_casp}(a)). Given a predicted pathway and an experimentally determined pathway, one strategy to perform their 3D structural comparison could be to generalize established structural similarity measures from comparing two native structures to comparing corresponding pairs of intermediates along the two pathways. This is precisely the strategy we have used to evaluate AlphaFold2-predicted co-translational intermediates in Section \ref{subsubsect:predict_cotrans}. While we have used TM-score (for reasons discussed in Supplementary Section \ref{sect:sup_tm_score}), additional 3D structural similarity measures can also be used, which are sometimes in agreement with and sometimes complementary to each other, depending on whether they are global vs. local, superposition-based vs. superposition-free, or considering all atoms vs. only selected subsets of atoms \cite{olechnovivc2019comparative}. Another quite natural strategy for 3D structural comparison of the two pathways could rely on dynamic PSNs. Namely, PSNs have been powerful in tasks such as protein structural classification \cite{faisal2017grafene,newaz2020network} as well as protein function prediction \cite{gligorijevic2021structure}. And of all PSN types, dynamic PSNs have shown the greatest accuracy \cite{newaz2022multi}. Dynamic PSNs were introduced exactly for the purpose of modeling 3D structures of intermediates along a co-translational pathway (Section \ref{subsubsect:co-trans_proxy}). So, each of a predicted pathway and an experimentally determined pathway can naturally be converted into its respective dynamic PSN. Then, the two pathways, i.e., their dynamic PSNs, can be compared using a wealth of  approaches for dynamic network comparison \cite{vijayan2017alignment,zitnik2024current}. This is especially true given the rapid advancement of artificial intelligence, including deep learning on graphs \cite{gligorijevic2021structure,zitnik2024current}, as well as geometric deep learning combined with network analyses \cite{morehead2024geometry,morehead2024geometry2}; these kinds of approaches have already shown strong performance in a range of structural biology tasks such as drug binding, protein–protein interaction prediction, and fold classification \cite{zitnik2024current}.

Second, when experimentally determined 3D structures of intermediates along a pathway are unavailable, one evaluation strategy could be to use quantitative data (e.g., folding rate constants or thermodynamic stability), which are more widely available than 3D structural data, especially for post-translational intermediates, like what Outeiral et al. (2022) \cite{Outeiral2022CurrentSP} and Huang et al. (2023, 2024) \cite{Huang2023PathfinderPF,zhao2023protein,zhao2024foldpathreader} did (Section \ref{subsect:comp_pathway}). Third, evaluation could involve performing in-house wet-lab experimental validation, as done by Wang et al. (2025) \cite{wang2025cotranslational}, per our discussion earlier in this section. 

In summary, the problem of protein folding dynamics is challenging, including due to data sparsity. However, such type of challenge is not without precedent. At the inception of PDB \cite{wwpdb2019protein} in 1976, only 13 structures were available. For its first decade, structural growth was slow, with only 6-32 additional structures being added to PDB each year (Supplementary Fig. \ref{fig:supp_pdb_n_casp}(a)). Yet over time, PDB has grown to include over 200,000 structures (Supplementary Fig. \ref{fig:supp_pdb_n_casp}(a)). In parallel, the launch of the Critical Assessment of protein Structure Prediction (CASP) competition \cite{yuan2025casp16} in 1994, with  27 teams and 186 predictive methods (Supplementary Fig. \ref{fig:supp_pdb_n_casp}(b)), sparked rapid progress in native structure prediction, culminating in tools like AlphaFold2 and an explosion of predicted 3D structures -- with over 214 million in the AlphaFold Protein Structure Database alone \cite{varadi2024alphafold}. This history suggests that major progress can begin with limited data. Such a growing trajectory, likely with a much faster timeline due to developments in artificial intelligence and machine learning, may be possible for data on 3D structures of folding intermediates if the community begins to systematically organize existing and generate new 3D structural data, however sparse; this has been exactly a key goal of our paper. Like the early PDB and CASP efforts on native structures, such a foundation could catalyze innovation in intermediate-specific experimental  technologies as well as computational prediction methods (importantly, including evaluation frameworks). A valuable and recent experimental attempt toward detecting co-translational intermediates is a modification to existing NMR spectroscopy, called $^{19}F$ NMR, which can detect and measure stable structural states accessed by a protein during co-translation \cite{chan2022ribosome}. 

Addressing the current impasse will require coordinated, closely intertwined efforts across wet lab experimental and computational communities. On the experimental side, new technologies are needed to capture transient 3D structures of intermediates with higher resolution and faster acquisition times. On the computational side, more efforts are needed to develop intermediate-specific prediction methods informed by curated experimental 3D structural data as well as kinetics or thermodynamics data. Additionally, we encourage the computational community to include analyses of computational complexity as well as results on running times and memory usage when evaluating new methods, as such key data are often overlooked in the current literature. Developing accurate and efficient computational approaches is only part of the endeavor; ensuring these methods are user-friendly and broadly accessible is equally critical for their widespread adoption. This includes offering intuitive interfaces and well-documented software, providing open-source implementations, and ensuring compatibility with platforms and workflows commonly used by the community.

\section*{Software and data availability}

Only existing, already published data and software are used in this paper, all of which are publicly available and appropriately referenced throughout the paper. For easiness and reproducibility, we provide a consolidated list of these, along with our predicted \underline{\textbf{3D}} structural data on ``proxy'' and AlphaFold2-predicted co-translational \underline{\textbf{p}}rotein \underline{\textbf{f}}olding \underline{\textbf{i}}ntermediates, at \url{https://github.com/Aywells/3Dpfi} or \url{https://www3.nd.edu/~cone/3Dpfi}.

\section*{Funding}

This work was supported by TM's appointment as a Frank M. Freimann Collegiate Associate Professor of Engineering at the University of Notre Dame, KN's funding under the Excellence Strategy of the Federal Government of Germany and the L\"{a}nder, and JC's NIH R01GM146340 award. Also, some ideas from this work were inspired by TM's past awards NSF CAREER CCF-1452795 and NIH 1R01GM120733. 

\bibliographystyle{abbrv} 

\bibliography{bibliography}

\newpage

\section*{Supplementary Information}\label{sect:supplement}
\beginsupplement
\section{\textcolor{black}{Supplementary data}}\label{sect:supporting_info}

    \subsection{Additional details on kinetics and thermodynamics data related to co-translational folding}\label{sect:sup_thermo}

        Here, we complement Section \ref{sect:introduction} in the main paper by discussing existing kinetics and thermodynamics data related to co-translational folding.

        For co-translational folding, we could not identify any organized database containing kinetics or thermodynamics data. Instead, we could identify some isolated studies that provide data of these types for a handful of proteins, as follows. Samelson et al. (2018) \cite{samelson2018kinetic} reported kinetics (co-translational folding rate) data for a protein called HaloTag,  Farias et al. (2018) \cite{farias2018effects} studied the thermodynamic stability of the S6 protein, and Kelkar et al. (2012) \cite{kelkar2012kinetic} captured kinetics data for intermediates of the protein mCherry at specific time points.

    \subsection{Additional information about the studies on experimentally determined 3D structures of post-translational intermediates}\label{sect:sup_post_studies}
    
        Here we complement Section \ref{subsect:exp_pathway} in the main paper by providing more details about the two studies on post-translational intermediates:
            
        Neudecker et al. (2012) \cite{neudecker2012structure} addressed the question of how folding intermediates contribute to amyloid fibril formation, a key process in many neurodegenerative disorders. While such intermediates have been implicated in aggregation, the structural mechanisms underlying this transition remain poorly understood. To investigate this, Neudecker et al. (2012) used NMR spectroscopy to determine the structure of a post-translational intermediate of the Fyn SH3 domain. Neudecker et al. (2012) found that the intermediate exhibited a disordered C-terminus, exposing an aggregation-prone $\beta$-strand. Two pathway intermediates are provided in the study: the first intermediate is an ``early-stage'' structure (PDB ID: 2L2P) and the second intermediate is the native structure (PDB ID: 2L25) of the SH3 domain.
    
        Zhou et al. (2008) \cite{zhou2008high} explored the folding mechanisms of ribonuclease H (RNase H) in order to enhance understanding of the principles governing protein folding dynamics. Utilizing multidimensional NMR, Zhou et al. (2008) identified a compact intermediate of RNase H that formed rapidly during folding, exhibiting a native-like core of helices. The study's findings supported a hierarchical folding mechanism, where secondary structures formed first, followed by tertiary contacts. Additionally, Zhou et al. (2008) findings indicated that the intermediate was well-defined and folded into its native structure more quickly because of this intermediate. Two post-translational intermediates are provided in the study: the first intermediate is an ``early-stage'' structure (PDB ID: 2RPI) and the second intermediate is the native structure (PDB ID: 1RIL) of the RNase H protein.

    \subsection{Additional information about the studies on experimentally determined 3D structures of co-translational intermediates}\label{sect:sup_co_studies}
    
        Here we complement Section \ref{subsect:exp_cotrans} in the main paper by providing more details about the four studies on co-translational intermediates:
    
        Agirrezabala et al. (2022) \cite{agirrezabala2022switch} investigated how early folding events are influenced by the ribosome -- and hence the shape of protein structures -- during translation. Specifically, Agirrezabala et al. (2022) observed the folding of a $\beta$-barrel protein (specifically a cold shock protein (CspA)) and reported its interactions with the ribosome during translation. Using cryo-EM, Agirrezabala et al. (2022) captured a co-translational intermediate of CspA that revealed an initial $\alpha$-helical conformation that formed within the ribosome's exit tunnel before transitioning into its native $\beta$-strand structure upon emergence. The study's findings emphasized the ribosome's role in shaping early folding events and suggested that co-translational folding may be a common feature of $\beta$-barrel proteins. Agirrezabala et al. (2022) provided the native structure of the CspA protein (PDB ID: 1MJC), three different conformations of the CspA protein when 27 amino acids have already been translated (PDB ID: 7NWW, 7OIF, 7OIG), and two different conformations of the CspA protein when 70 amino acids have already been translated (PDB ID: 7OT5, 7OII).
    
        Hanazono et al. (2018) \cite{hanazono2018co} explored the co-translational folding of nascent polypeptide chains, specifically focusing on the $\lambda$ repressor N-terminal domain, a small $\alpha$-helical protein, to uncover the atomic-level details of this process. Using circular dichroism (CD) spectroscopy, Hanazono et al. (2018) examined intermediate-length variants of the $\lambda$ repressor to capture structural intermediates during co-translational folding. The study found that partial helices formed within the ribosome's exit tunnel, with increasing chain length leading to progressive stabilization of secondary and tertiary structure. This suggested a stepwise folding process where local $\alpha$-helical segments formed early and guided subsequent structural changes. Their results highlighted how the ribosome constrains and influences protein stability and function. Hanazono et al. (2018) reported the native structure of the $\lambda$ repressor (PDB ID: 5ZCA), and the structures of two different co-translational folding intermediates; the first of the two captures the structure of the $\lambda$ repressor when 20 amino acids have already been translated (PDB ID: 3WOA) and the second captures the structure of the $\lambda$ repressor when 45 amino acids have already been translated (PDB ID: 1LMB).
    
        Hanazono et al. (2016) \cite{hanazono2016structural} investigated the co-translational folding of nascent proteins, which are affected by the rate of translation by the ribosome. While fully translated proteins are capable of achieving their native conformation, nascent protein structures primarily begin folding co-translationally at their N-terminal regions; however, the transient structures of intermediates involved in this process remain poorly understood. Hanazono et al. (2016) focused on the early folding events of the WW domain, a small $\beta$-sheet protein, during translation. Using CD spectroscopy, N-terminal fragments of the WW domain were examined to determine how partial sequences adopted structure before arriving at the native structure. Hanazono et al. (2016) showed that isolated N-terminal fragments lack stable $\beta$-sheet formation, suggesting that the WW domain required a sufficiently long polypeptide chain before $\beta$-strands can properly fold. This contrasts with $\alpha$-helical structures, where partial helices formed early. Hanazono et al. (2016) findings highlighted a cooperative folding mechanism for $\beta$-sheet proteins, where both chain length and long-range molecular interactions played a key role in stabilizing the WW domain's structure. The study reported the native structure of the WW domain (PDB ID: 5B3Z), and the structures of three different N-terminal fragments of the WW domain: the first of the three captures the structure of the WW domain when 11 amino acids have already been translated (PDB ID: 3WOA), the second of the three captures the structure of the WW domain when 17 amino acids have already been translated (PDB ID: 5BMY), and the last of the three captures the structure of the WW domain when 19 amino acids have already been translated (PDB ID: 5B3Y).
    
        Cabrita et al. (2016) \cite{cabrita2016structural}, using NMR spectroscopy, analyzed  folding of FLN5 in isolation (i.e., its post-translational folding) vs. folding of FLN5 in the presence of the ribosome (i.e., FLN5's co-translational folding). Specifically:
        \vspace{-0.35cm}
        \begin{itemize}
        \item \textcolor{black}{Regarding the former: isolated FLN5 was observed to fold spontaneously, corresponding to FLN5's native 3D structure; we show this structure, reported by Cabrita et al. (2016), as intermediate 1 in Fig. \ref{fig:cotrans_data} of the main paper (PDB ID: 1QFH -- now updated to 6G4A).} 
        \vspace{-0.25cm}
        \item \textcolor{black}{Regarding the latter:     
        Cabrita et al. (2016) examined how much time needed to pass between the entire FLN5 being out of the ribosome tunnel before obtaining its native  structure (i.e., a structure closely matching the native structure of isolated FLN5). To be able to measure this, Cabrita et al. (2016) attached to FLN5 an increasingly longer portion of the sequence of FLN6, with FLN5 being at the N-terminal of the combined sequence, and FLN6 being at its C-terminal. Then, appending a shorter subsequence of FLN6 to (the full-sequence of) FLN5 meant less time had passed since FLN5 was translated by the ribosome, while appending a longer subsequence of FLN6 to (the full-sequence of) FLN5 meant more time had passed since FLN5 was translated by the ribosome. What Cabrita et al. (2016) then found was that FLN5 could obtain the native structure in the presence of the ribosome only when a longer (full) subsequence of FLN6 was appended to it. In other words, the complete sequence of FLN5 had to emerge well beyond the ribosome tunnel before it could  acquire its native structure (while we note again that FLN5 in isolation folded spontaneously). Cabrita et al. (2016) concluded this to be  evidence that the ribosome modulates the co-translational folding process of FLN5. Cabrita et al. (2016) reported three distinct 3D structural confirmations of the combined FLN5+FLN6 sequence, corresponding to intermediate 2 -- conformations a, b, and c -- in Fig. \ref{fig:cotrans_data} in the main paper (PDB ID: 2N62).  Note that per the above discussion of Cabrita et al. (2016) results, (only) the FLN5 portion of these three 3D structural confirmations of FLN5+FLN6 (i.e., the green part of confirmations 2a, 2b, and 2c in the ``Cabrita et al., (2016)'' portion of Fig. \ref{fig:cotrans_data} in the main paper) is a (close) structural match to the isolated native 3D structure of FLN5 (i.e., intermediate 1 in the ``Cabrita et al., (2016)'' portion of Fig. \ref{fig:cotrans_data} in the main paper).}
        \end{itemize}
    
\section{\textcolor{black}{Supplementary methods}}\label{sect:supporting_info}

    \subsection{Additional details about our use of TM-score}\label{sect:sup_tm_score}
    
        Here we provide details on our use of TM-score \cite{xu2010significant, zhang2004scoring}.

        TM-score is a widely used quantitative measure for assessing the level of similarity between two 3D structures. It is a \textit{global} measure, meaning that it evaluates the overall agreement of entire folds between two 3D structures, even if some of their peripheral regions might differ \cite{zhang2004scoring}.  
        
        In contrast to global 3D structural similarity measures, \textit{local} measures  assess structural similarity at a finer spatial scale; these measures typically involve calculating the spatial distance deviations of pairs of residues (typically $C\alpha$ atoms) within a \textit{local neighborhood}, and then aggregating scores over only short-range residue interactions. As such, local structural similarity measures are sensitive to local distortions, making them useful for evaluating the accuracy of smaller structural regions regardless of the global fold \cite{mariani2013lddt}. 

        It was recently shown that global measures of 3D structural similarity -- including TM-score and Global Distance Test (GDT) -- are ``extremely highly correlated'' \cite{olechnovivc2019comparative}, which supports the use of TM-score as a representative global measure. The same study found that local measures -- including local Distance Difference Test (lDDT), Recall, Precision and F-measure (RPF), and Contact Area Difference (CAD) -- are also ``highly correlated'' with each other \cite{olechnovivc2019comparative}. Further, that study found that the local similarity scores are often in agreement with global ones, as well as that the methods for predicting 3D structures (referred to as ``models'' in that study) that are selected as the best with respect to local measures (including lDDT and RFP) are typically also among the best methods selected according to the global measures (including TM-score and GDT) \cite{olechnovivc2019comparative}. Yet, the two groups of measures, global vs. local, as well as individual measures within each group, often have (dis)advantages \cite{olechnovivc2019comparative}. 

        Since one aim of our study is to determine whether two compared 3D structures (as described in cases (i) and (ii) below) have the same \textit{overall} fold, we use TM-score, which is global, as our primary measurement for assessing 3D structural similarity. As mentioned above, local structural similarity measures might provide valuable complementary information in certain aspects, particularly when the global fold is preserved. However, we argue that if the global fold is not preserved between the compared 3D structures -- as we have observed in our analyses of non-native folding intermediates, where TM-scores are almost always below 0.5 -- identifying specific regions of local deviation would offer limited structural insights, as the underlying folds are fundamentally distinct. Given this, and given the above discussion of frequent agreement/consistency as well (rather than just complementarity) between results of global and local measures, we believe that the use of TM-score as our measure of choice in our study is justified as  well as sufficient.

        We use the TM-score \cite{zhang2004scoring} software available through the Zhang lab's web server (\hyperlink{https://zhanggroup.org/TM-score/}{https://zhanggroup.org/TM-score/}). We compare 3D structures of two intermediates (i.e., their sequences) in terms of TM-score as follows. We either compare (i) the experimentally determined 3D structure of a given sequence to the experimentally determined 3D structure of the same sequence \emph{at a different time} (Section \ref{sect:exp_data} in the main paper), or (ii) the \emph{experimentally determined vs. predicted} 3D structures a given sequence (Section \ref{subsect:comp_cotrans} in the main paper).  Because there exist some (minor) discrepancies between the deposited and modeled sequences of a given intermediate (the  hatched boxes in Fig. \ref{fig:cotrans_data} in the main paper) for several of the 15 considered conformation of the 10 considered intermediates (all but the two red conformations of intermediates in Fig. \ref{fig:cotrans_data} in the main paper), there also may exist discrepancies between the modeled sequences of two compared intermediates. If so, we handle these discrepancies when computing TM-scores between two modeled sequences as follows. For case (i) above, we first identify and extract the longest common contiguous subsequence (LCCS) shared between the experimentally determined compared modeled sequences. Each of the two resulting LCCS modeled sequences is then converted into PDB format and uploaded to the TM-score web server mentioned above using the ``Structure 1'' and ``Structure 2'' input fields. Then, the web server returns the corresponding TM-score. For case (ii) above, recall that we predict a 3D structure for an intermediate by inputting the \emph{deposited} sequence of the intermediate into AlphaFold2 (Section \ref{subsubsect:predict_cotrans} in the main paper). Then, when we compare a predicted 3D structure and its experimentally determined counterpart. Here, we first extract the LCCS between the deposited sequence of the predicted 3D structure and the modeled sequence of its corresponding experimentally determined 3D structure. Second, the two resulting 3D structures corresponding to the two LCCSs are then converted into PDB format and uploaded to the TM-score web server mentioned above using the ``Structure 1'' and ``Structure 2'' input fields. Then, the web server returns the corresponding TM-score. Note that in either of cases (i) and (ii) above, we do not need to normalize a given TM-score with respect to the length of either of the two input 3D structures, because we compare only the LCCS shared between the pair; i.e., the two structures inputted into the TM-score web server always have the exact same sequence. 
        
        To reemphasize, this methodology is applied uniformly across all analyses presented in Sections \ref{sect:exp_data} and \ref{subsect:comp_cotrans} of the main paper where TM-score is employed.

        TM-scores range from 0 to 1, with higher values indicating greater structural similarity between protein 3D structures. In more detail \cite{xu2010significant}: A TM-score below 0.17 indicates a random-like structural similarity, i.e. that the probability of finding a TM-score this low from randomly chosen structural pairs is close to 1. A TM-score higher than 0.3 indicates a significantly high structural similarity, i.e. that the probability of finding such a score from random structural pairs is very low. A TM-score of 0.5 is generally considered a key threshold, as a TM-score higher than 0.5 suggests that the compared 3D structures have the same overall fold or similar topology according to structural classifications such as CATH \cite{greene2007cath} and SCOPe \cite{fox2014scope}.

    \subsection{Additional details on the evaluation strategies used in the Outeiral et al. (2022) study}\label{sect:sup_comp_posttrans_deates}

        Here we complement Section \ref{subsect:comp_cotrans} in the main paper with more details on the evaluation from the study by Outeral et al. (2022) \cite{Outeiral2022CurrentSP}.
        
        Recall that Outeral et al. (2022) investigated whether 3D structures of \textit{post-translational} intermediates could be accurately predicted with existing methods designed for predicting native 3D structures \cite{Outeiral2022CurrentSP}. They asked this question for eight such methods: AlphaFold2 \cite{Jumper2021HighlyAP}, RoseTTAFold \cite{baek2021accurate}, trRosetta \cite{yang2020improved}, RaptorX \cite{peng2011raptorx}, DMPfold \cite{greener2019deep}, EVfold \cite{marks2011protein}, SAINT2 \cite{de2018sequential}, and Rosetta \cite{schaap2001rosetta}. More specifically, for each of these methods,  Outeral et al. (2022) asked whether a given method could predict post-translational pathways that are (i) predictive of a protein's folding kinetics class (e.g. two-state or multi-state), and (ii) correlate with experimentally measured folding rate constants. More details about these two evaluation tasks are as follows.

        For the first evaluation task, the methods' predicted pathways were used to classify whether the given protein folds through two-state or multi-state kinetics, in both an unsupervised and supervised manner. \textcolor{black}{Multiple measures of method performance accuracy were used, including the accuracy, F1-score, and area under the receiver-operating characteristic (AUROC) curve. In the text, we summarize the results with respect to AUROC as a representative measure for illustration purposes (where AUROC  score of 0.50 indicates random performance, and the higher AUROC score, the better the given method). The AUROC results (for 10 pathways per method per protein, except for AlphaFold2 with one pathway per protein) are as follows. All methods achieved statistically significant yet quite modest performance, with AUROC scores between 0.56 and 0.675. The actual protein structure prediction methods were compared to a trivial baseline, namely a simple linear classifier based solely on chain length. Surprisingly, this simple baseline outperformed all structure prediction methods, with AUROC score of 0.739. Results were qualitatively similar for the other performance accuracy measures and when predicting 200 pathways per method per protein. Full results for this evaluation task are shown in Supplementary Table \ref{tab:outerial_study_results}. }   
        A key conclusion from this analysis was that this sequence-agnostic baseline surpassed all structure-based methods in predicting folding kinetics, indicating that the predictive signal captured by current structure prediction tools is weak.

        For the second evaluation, the methods were evaluated based on how well their predicted pathways could capture the folding rate constants of proteins that undergo two-state folding kinetics. To test this, Outeral et al. (2022) selected the 79 proteins that had experimentally determined folding rate constants and had at least one pathway classified as two-state. Within these, only the predicted pathways classified as two-state were retained. For each retained pathway, Outeral et al. (2022) determined the relative frame in which folding occurred, defined as the point of maximal increase in native contacts. Outeral et al. (2022) then examined whether this position along the pathway correlated with the experimentally measured folding rate constants. Outeral et al. (2022) compared these correlations of the structure prediction methods to correlations obtained using two baseline methods -- average contact order and protein chain length -- both being known predictors of folding rates. The results showed that chain length had the strongest (and correct-sign) correlation, outperforming contact order and any structure prediction method. Most structure prediction methods showed weak or insignificant correlations, and in some cases even the wrong sign. Only AlphaFold2 and RoseTTAFold showed modest, correctly signed correlations, hinting at a limited signal. Outeral et al. (2022) findings suggested that while post-translational folding pathways predicted by structure prediction methods may resemble real post-translational pathways in some aspects, they fail to meaningfully correlate with experimental folding rate data. The original results for the second evaluation task by Outeral et al. (2022) can be found in Supplementary Fig. \ref{fig:outerial_study_results_2}.
        
\section{\textcolor{black}{Supplementary results}}\label{sect:supporting_info}

    \subsection{Results on structural similarities between conformations of intermediates for the same protein, for experimental co-translational folding data }\label{sect:sup_exp_cotrans_deates_2}
    
        Here we complement Section \ref{subsect:exp_cotrans} in the main paper by reporting key observations and methodological details about structural similarities in terms of TM-scores between all possible pairs of (partial as well as full) conformations of intermediates of the same protein, for experimental co-translational folding data.  Relevant results are shown in Supplementary Fig. \ref{fig:co_trans_n_TMscore_mtx}.
        
        First, we focus on different conformations for the same intermediate, which only the studies by Agirrezabala et al (2022) and Cabrita et al. (2016) have. In other words, we focus on 1a vs. 1b vs. 1c as well as 2a vs. 2c in the former, and 2a vs. 2b vs. 2c in the latter. All of the corresponding TM-scores are below 0.5, indicating that no two conformations of the same intermediate have the same fold. 
        
        Second, we focus on the conformational change of a sequence of an intermediate gradually over time, which all four studies have. For example, we focus on the conformation of a green sequence in one intermediate vs. the conformation of the same green sequence in the next intermediate (for the same protein), or on the conformation of a green+orange sequence in one intermediate  vs. the conformation of the same green+orange sequence in the next intermediate  (for the same protein). Overall, we observe quite a large conformational change of the same sequence over time in the presence of additional amino acids being translated and added to the 3D structure, most often resulting in a changed fold of the given sequence during translation. In more detail, for the protein from Agirrezabala et al (2022), we observe the fold change in all comparisons (all TM-scores are 0.33 or lower). For the protein from Hanazono et al (2018), we also observe the fold change in all comparisons (all TM-scores are 0.43 or lower). For the protein from  Hanazono et al (2016), three of the four comparisons indicate a fold change (TM-scores of 0.46 or lower), and in the remaining case, there is still a large structural change although not necessarily a fold change (TM-score of 0.58). Finally, for the protein from Cabrita et al. (2016), we see the highest TM-scores, meaning the least amount of conformational change, with TM-scores between 0.77 and 0.82. The higher TM-scores for the Cabrita et al. (2016) protein compared to the other three proteins/studies are expected given the unique nature of the Cabrita et al. (2016) study, as described in the last paragraph of Supplementary Section \ref{sect:sup_co_studies}, i.e., because intermediate 1 is a native structure of isolated (post-translationally folded) FLN5, and (only) the green portions of intermediates 2a, 2b,  and 2c are native structures of co-translationally folded FLN5.
        
        Third, in order to evaluate the effect of time passed, we compare conformations corresponding to the same sequence of an intermediate at closer vs. more distant times. Specifically, for the only two studies that allow for this comparison (because they have more than two distinct intermediates in Supplementary Fig. \ref{fig:co_trans_n_TMscore_mtx}) -- Hanazono et al (2018) and Hanazono et al (2016) -- we compare conformational change of the green sequence from time 1 to time 2, and from time 2 to time 3, against its change from time 1 to time 3. For the protein from Hanazono et al (2018), we find the following. The green sequence in the first intermediate vs. the green sequence in the second intermediate, as well as the green sequence in the second intermediate vs. the green sequence in the third intermediate (i.e. the two pairs of time-closest intermediates), show a high level of conformational changes (TM-scores of 0.43 and 0.42, respectively). Yet, the green sequence in the first intermediate vs. the green sequence in the third intermediate (i.e. the pair of most time-distant intermediates) show an even greater level of conformational change (TM-score of 0.22). Similarly, in Hanazono et al (2016), the green sequence in the first intermediate vs. the green sequence in the second intermediate, as well as the green sequence in the second intermediate vs. the green sequence in the third intermediate, show some conformational changes (TM-scores of 0.45 and 0.58, respectively). The green sequence in the first intermediate vs. the green sequence in the third intermediate show an even larger conformational change (TM-score of 0.43).

    \subsection{Per-study results of AlphaFold2 predictions of co-translational intermediates }\label{sect:sup_comp_posttrans_deates_4}

        Here we complement Section \ref{subsubsect:predict_cotrans} in the main paper with \emph{per-study} results related to our analysis of AlphaFold2-predicted co-translational intermediates. Relevant results are shown in Fig. \ref{fig:AF2_n_TMscore} in the main paper.

        For the Hanazono et al. (2018) and Hanazono et al. (2016) studies, a common trend in TM-scores is apparent; in each of these studies, AlphaFold2's predicted structure of the last intermediate is highly structurally similar (with TM-scores greater than $0.80$) to the last modeled sequence (i.e. intermediate 3 in Hanazono et al. (2018), and intermediate 4 in Hanazono et al. (2016)). This is because the modeled sequence of the last intermediate in each of these studies is the native structure (as reported in it's respective study). However, most of AlphaFold2's predicted structures of earlier intermediates do not have the same fold as their experimental counterparts (with TM-scores lower than $0.50$, with the exception of intermediate 2 in Hanazono et al. (2018) with TM-score $=0.53$).

        For the Agirrezabala et al. (2022) study a similar trend is also observed, but, with different insights between different conformations of intermediates; for conformation 2c, AlphaFold2's predicted structure is highly structurally similar to it (with TM-score $=0.95$), because the modeled sequence of this conformation is the native structure. In contrast, for conformation 2a -- an experimentally observed non-native structure -- AlphaFold2's predicted structure is highly dissimilar to it, close to random (with TM-score $=0.22$). Although these two conformations have the same deposited sequences, when given a modeled sequence as input, AlphaFold2's predicted structure is biased towards predicting a native structure with high confidence, rather than a structure of an intermediate. Furthermore, Alphafold2 cannot correctly predict any of the first intermediate conformations (i.e. conformations 1a, 1b, and 1c, with TM-scores less than or equal to $0.17$).

        At first glance, for the Cabrita et al. (2022) study, results show a different pattern, with AlphaFold2's predicted structure of the first intermediate being highly structurally  similar to the modeled sequence of the first intermediate (TM-score $=0.82$), and AlphaFold2's predicted structures of the conformations of the second intermediate not having the same fold for most of the conformations (with TM-scores ranging between $0.45$ and $0.53$). However, the first intermediate is the native structure (as reported in the Cabrita et al. (2022) study), and the conformations of the second intermediate are non-native structures \textcolor{black}{(even though their green portions in Fig. \ref{fig:cotrans_data} in the main paper are native structures, per our discussion in the last paragraph of Supplementary Section \ref{sect:sup_co_studies})}. While AlphaFold2 accurately predicts the native structure (i.e. intermediate 1), it is unable to predict any conformations of the second intermediate with high structural similarity to experimental data.

    \subsection{Additional details on structural similarity between experimentally determined co-translational intermediates and their corresponding ``proxy'' vs. AlphaFold2-predicted co-translational intermediates}\label{sect:sup_comp_posttrans_deates_5}

        Here we complement Section \ref{subsubsect:co-trans_proxy} in the main paper with more details about the analysis of 3D structural similarity between experimentally determined intermediates and their corresponding ``proxy'' vs. AlphaFold2-predicted intermediates. 

        In this analysis, of the four considered studies/proteins, we focus on those studies that meet two key criteria. First, for a given protein, its native structure must correspond to the intermediate with the longest sequence out of all intermediates. This ensures that the native structure corresponds to the full-length conformation, which is necessary for generating ``proxy'' intermediates by extracting substructures from the native fold. The protein from the Cabrita et al. (2022)3 \cite{cabrita2016structural} study is excluded from further analysis because it violates this criterion: none of the three conformations of the intermediate with the longest sequence (conformations a, b, and c of intermediate 2) are the native structure; the native structure corresponds to the shortest-sequence intermediate (intermediate 1); for more details, see the last paragraph of Supplementary Section \ref{sect:sup_co_studies}. Second, for a given protein, only a single conformation should be available for any intermediate. This constraint is desirable because if multiple different conformations exist for the same sequence, AlphaFold2 would incorrectly predict the same 3D structure for all of those confirmations, given their shared sequence. The protein from the Agirrezabala et al. (2022) \cite{agirrezabala2022switch} study fails to meet this criterion, as it has three conformations for its intermediate 1 and two conformations for its intermediate 2; also, the protein from the Cabrita et al. (2022) study fails this criterion too, as it has three confirmations for its intermediate 2. Only the proteins from the Hanazono et al. (2018) \cite{hanazono2018co} and Hanazono et al. (2016) \cite{hanazono2016structural} studies satisfy both conditions and are thus considered in the rest of this analysis.

        We infer AlphaFold2-predicted intermediates for the proteins in the two considered studies as follows: For the protein in the Hanazono et al. (2018) study, we input the experimentally determined sequence of intermediate 1 into AlphaFold2 to obtain the corresponding AlphaFold2-predicted intermediate; similarly, we input the experimentally determined sequence of intermediate 2 into AlphaFold2 to obtain the corresponding AlphaFold2-predicted intermediate. For the protein in the Hanazono et al. (2016) study, we follow the same procedure using the experimentally determined sequences of intermediates 1 and 2  as inputs into AlphaFold2 to obtain their corresponding AlphaFold2-predicted intermediates. Note that we do not generate AlphaFold2-predicted intermediates for the native structures themselves for the same reason as stated above. In total, two AlphaFold2-predicted intermediates are generated for each protein from each of the two considered studies, and their structural similarities to their respective experimentally determined counterparts are computed using TM-score (these scores were originally computed in Fig. \ref{fig:AF2_n_TMscore} in the main paper, but are shown again for the purpose of this analysis in  Table \ref{tab:supp_af2_proxy_compare} in the main paper).

\clearpage

\section{Supporting figures and tables for the main paper}\label{sect:supporting_figs}

\clearpage

    \begin{figure*}
        \begin{center}
        \includegraphics[width=0.95\textwidth,trim= 0.5cm 5cm 0.5cm 1.5cm]{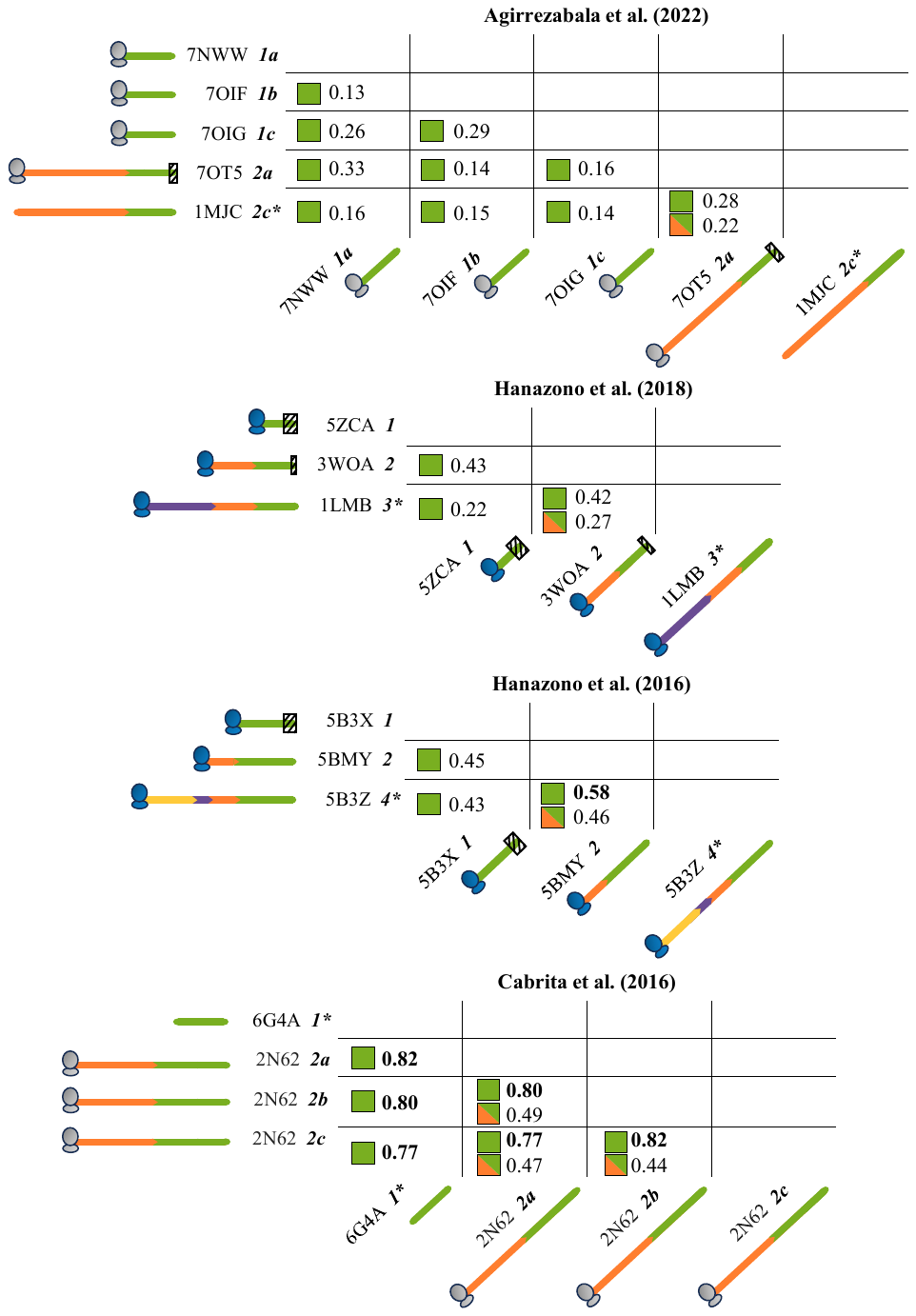}
        \end{center}
        \caption{\textcolor{black}{Pairwise structural similarities in terms of TM-score between different combinations of (the conformations of) the co-translational intermediates from the same protein/study.  The four proteins/studies considered are the same as in Fig. \ref{fig:cotrans_data} in the main paper.  Their conformations of the intermediates are shown again in the matrix rows/columns of this figure for easiness, with the exception of the conformations with PDB IDs 70II and 5B3Y, which are excluded from any analyses, per the discussion in Fig. \ref{fig:cotrans_data} in the main paper; all colors and hashed regions in the current figure match those in Fig. \ref{fig:cotrans_data} in the main paper. There are four matrices of pairwise similarities corresponding to the four proteins/studies, one matrix per protein. In each matrix, TM-scores are shown between all possible pairs of conformations of the intermediates when accounting for their modeled sequence regions. A green box  next to a score indicates that only the green regions of the pair of modeled sequences are compared; an orange and green box next to a score indicates that the orange and green regions combined of the pair of modeled sequences are compared. All TM-scores with values higher than  $0.50$ are bolded, corresponding to structures that have the same overall fold \cite{zhang2004scoring}. Only the bottom triangle of the  matrix is filled with TM-scores, because the matrix is symmetric.}}
        \label{fig:co_trans_n_TMscore_mtx}
    \end{figure*}

\clearpage

    \begin{figure*}
        \begin{center}
        \includegraphics[width=\textwidth,trim= 0.25cm 10.2cm 4.75cm 0.5cm]{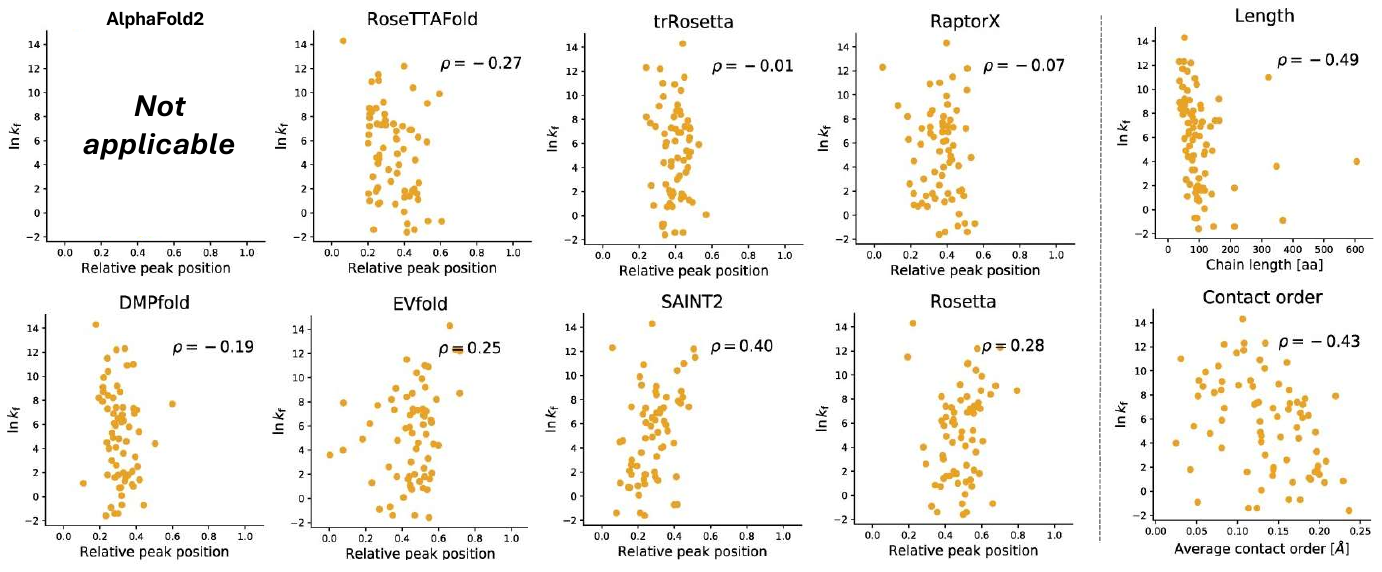}
        \end{center}
        \caption{\textcolor{black}{The figure and results from the  \cite{Outeiral2022CurrentSP} study on the performance of their eight considered protein structure prediction methods and the chain length (``Length'') method as a basic benchmark approach, \underline{in their second considered task} of examining whether the predicted post-translational pathways correlate with experimentally measured folding rate constants. As stated in the original publication \cite{Outeiral2022CurrentSP}: The illustrated analysis was used to compute the Spearman correlation coefficient \textcolor{black}{$\rho$} between the folding rate constant and folding events in simulated trajectories (i.e., predicted pathways) of the seven considered structure prediction methods, as well as the length of the protein chain and the average contact order of the native structure. Every point represents the average over the maximum number of decoys possible (200 decoys for RoseTTAFold, trRosetta, RaptorX, DMPfold and EVfold; and 10 decoys for SAINT2 and Rosetta). The panel for AlphaFold2 was not included by \cite{Outeiral2022CurrentSP} into this figure, because for AlphaFold2 only one decoy was available; however \cite{Outeiral2022CurrentSP} reported for AlphaFold2 that the correlation coefficient between the relative position of the folding event and the logarithm of the $k_f$ was -0.23'', which is of the same order and same sign as RoseTTAFold. In the figure, the best-performing method is Length, as it has the highest absolute-value \emph{and} correctly signed (negative) correlation coefficient of -0.49.}}
        \label{fig:outerial_study_results_2}
    \end{figure*}

\clearpage

\clearpage

    \begin{figure*}
        \begin{center}
                \includegraphics[
                    width=0.95\textwidth,
                    trim= 0.25cm 7.5cm 1cm 0cm,
                    clip
                ]{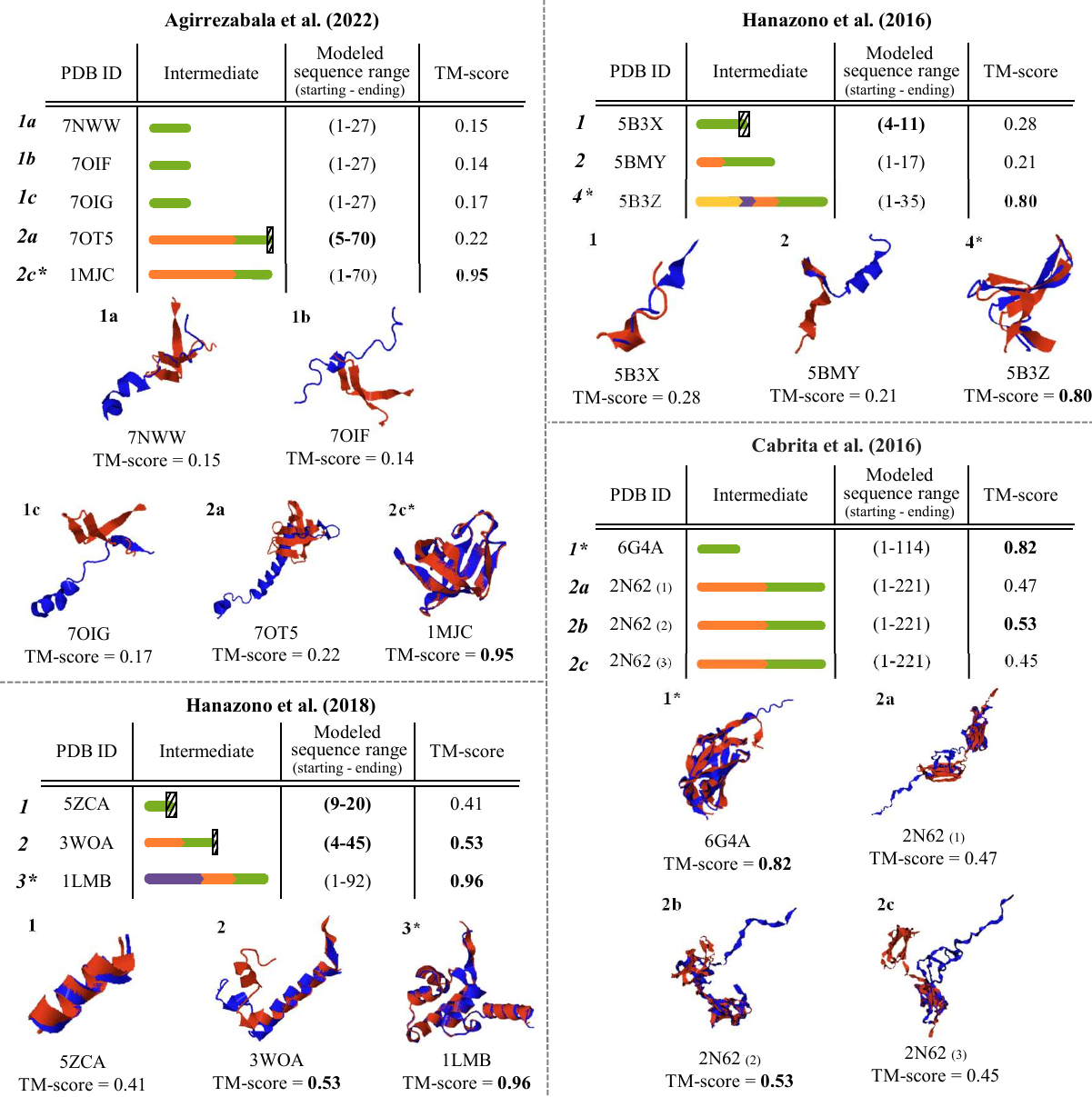}%
        \end{center}
        \caption{Structural similarities in terms of TM-scores between Alphafold2-predicted structures and experimentally determined structures for the  15 considered conformations of the 10 co-translational intermediates (i.e. for all but the two red ones in Fig. \ref{fig:cotrans_data} in the main paper). For each of the four studies, there is a corresponding table; the first three table columns are already explained in Fig. \ref{fig:cotrans_data} in the main paper (except that here we do not show or analyze an attached entity at the C-terminus). The fourth table column reports the TM-score of the highest-ranked predicted structure from AlphaFold2 (i.e. rank 1 from Supplementary Fig. \ref{fig:supp_af2_tm_score_rank2-5}). All TM-scores with values higher than  $0.50$ are bolded, corresponding to structures that have the same overall fold. For each TM-score, corresponding to a conformation of an intermediate, a 3D visualization of the AlphaFold2 predicted structure (blue) is overlaid with a 3D visualization of the experimental 3D structure corresponding to the modeled sequence (red).}
        \label{fig:supp_AF2_n_TMscore}
    \end{figure*}

\clearpage

    \begin{figure*}
        \begin{center}
        \includegraphics[width=1\textwidth,trim= 0cm 0.5cm 0cm 0cm]{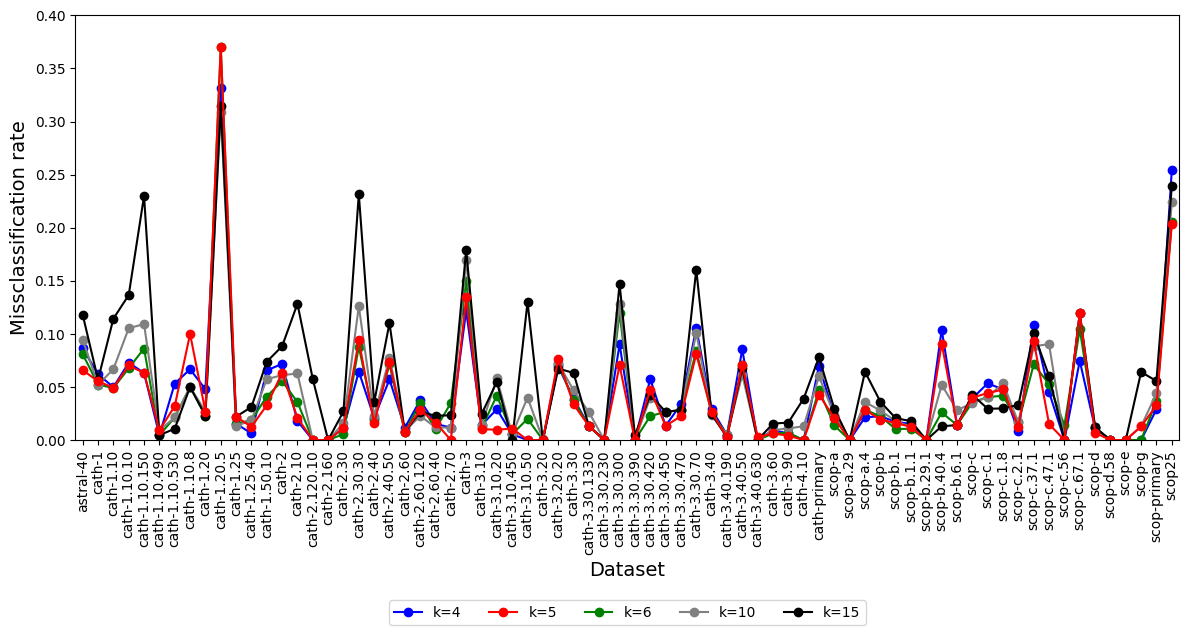}
        \end{center}
        \caption{The effect of constructing dynamic PSNs with \textcolor{black}{different values for $k$. As briefly mentioned in Section \ref{subsubsect:co-trans_proxy} in the main paper, dynamic PSNs were evaluated in the protein structure classification (PSC) task, specifically in the task} of classifying $\sim$44,000 protein domains with respect to their CATH \cite{greene2007cath} and SCOPe \cite{fox2014scope} structural classes. In the original dynamic PSN study \cite{newaz2022multi}, the $\sim$44,000 protein domains were organized into 72 protein domain datasets ($x$-axis). Also in the original study, dynamic PSNs constructed when $k=5$ (red in the figure; Section \ref{subsubsect:co-trans_proxy} in the main paper) were evaluated in the task of PSC with respect to misclassification rate (\textit{y}-axis, where lower is better). As an original contribution to our current study, we evaluate the effect on PSC misclassification rate of constructing dynamic PSNs with alternative values for $k$ (specifically $k = 4, 6, 10, 15$, corresponding to blue, green, grey, and black in the figure, respectively). We find that out of the 72 datasets, $k=5$ performs the best (i.e. has the lowest misclassification rate) on 34 datasets, and is within 1\% misclassification rate of the best other value(s) of $k$ on an additional 24 datasets. In other words, it is (close to) tied with, or better than, all other considered values of $k$ on $(34+24)/72=80.6\%$ of all analyzed datasets. These findings confirm that $k=5$ is quite a meaningful parameter choice for constructing dynamic PSNs.}
        \label{fig:supp_k_val_alt}
    \end{figure*}

\clearpage

\clearpage

    \begin{table*}
        \begin{center}
        \includegraphics[width=\textwidth,trim= 0cm 9cm 3cm 0cm]{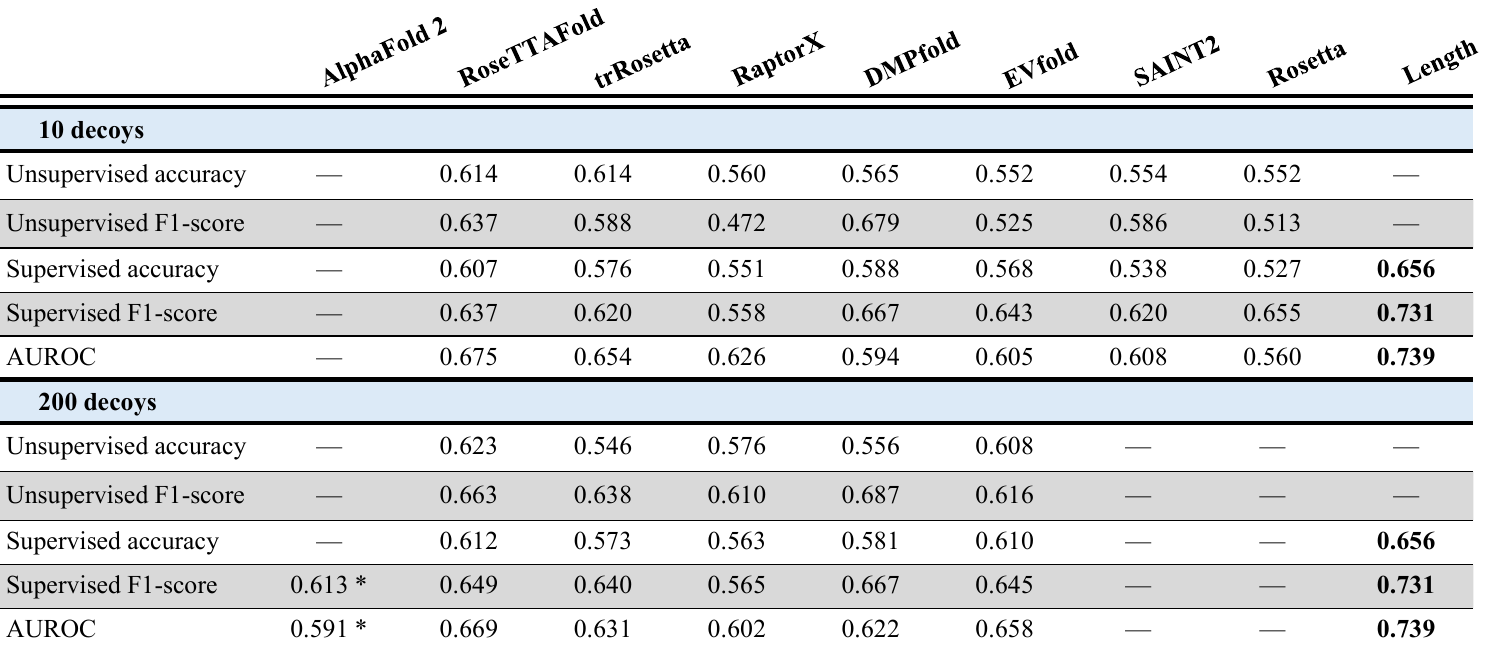}
        \end{center}
        \caption{\textcolor{black}{Results from the Outeiral et al. (2022) \cite{Outeiral2022CurrentSP} study on the performance of their eight considered protein structure prediction methods and the chain length (``Length'') method as a basic benchmark approach, \underline{in their first considered task} of examining whether the predicted post-translational pathways predict proteins' folding kinetics classes. The two blue rows correspond to the number of predicted post-translational pathways (i.e., decoys) generated per protein by each method. The remaining rows contain the performance results for the supervised and unsupervised aspect of the considered task, with respect to three evaluation measures (accuracy, F1-score, and AUROC).  Columns represent the $8+1$ considered methods. In a given row, the bolded value indicates the best performing method. Performance scores that have an asterisk indicate that the score was computed with respect to only one decoy, rather than 10 or 200 decoys.}}
        \label{tab:outerial_study_results}
    \end{table*}

\end{document}